\documentclass[a4paper,11pt]{article}
\usepackage[utf8]{inputenc}

\usepackage{geometry}
\usepackage{setspace}
\usepackage{authblk}

\usepackage{array, threeparttable, booktabs,caption}
\usepackage{times}
\usepackage{enumitem}
\usepackage{hyperref}
\hypersetup{colorlinks,allcolors=black}
\usepackage{tabularx}
\usepackage{graphicx}
\usepackage{adjustbox}
\newcommand{\RNum}[1]{\uppercase\expandafter{\romannumeral #1\relax}}
\newcommand{\beginsupplement}{%
        \setcounter{section}{0}
        \renewcommand{\thesection}{S\arabic{section}}
        \setcounter{table}{0}
        \renewcommand{\thetable}{S\arabic{table}}%
        \setcounter{figure}{0}
        \renewcommand{\thefigure}{S\arabic{figure}}%
     }

\begin{document}

\title{\textbf{Ranking mobility and impact inequality in early academic careers}}
\author[1]{Ye Sun}
\author[1,2,3]{Fabio Caccioli}
\author[1,2,*]{Giacomo Livan}

\affil[1]{Department of Computer Science, University College London, 66-72 Gower Street, London WC1E 6EA, United Kingdom}
\affil[2]{London School of Economics and Political Science, Systemic Risk Centre, London WC2A 2AE, United Kingdom}
\affil[3]{London Mathematical Laboratory, United Kingdom}

\affil[*]{To whom correspondence and requests for materials should be addressed. E-mail: g.livan@ucl.ac.uk}
\date{}

\maketitle

\begin{abstract}
    How difficult is it for an early career academic to climb the ranks of their discipline? We tackle this question with a comprehensive bibliometric analysis of 57 disciplines, examining the publications of more than 5 million authors whose careers started between 1986 and 2008. We calibrate a simple random walk model over historical data of ranking mobility, which we use to (1) identify which strata of academic impact rankings are the most/least mobile and (2) study the temporal evolution of mobility. By focusing our analysis on cohorts of authors starting their careers in the same year, we find that ranking mobility is remarkably low for the top and bottom-ranked authors, and that this excess of stability persists throughout the entire period of our analysis. We further observe that mobility of impact rankings has increased over time, and that such rise has been accompanied by a decline of impact inequality, which is consistent with the negative correlation that we observe between such two quantities. These findings provide clarity on the opportunities of new scholars entering the academic community, with implications for academic policymaking.
\end{abstract}

\section*{Introduction}
Recognition and rewards in modern academia are highly stratified~\cite{cole1974social}. The citations received by authors and their work~\cite{aksnes2009researchers,ruiz2014skewness,ruiz2018individual}, the amount of funding allocated to research projects~\cite{ma2015anatomy,bloch2015size}, and the number of  prizes awarded to scientists~\cite{ma2018scientific} are all very unevenly distributed quantities. 

The Matthew effect explains uneven outcomes in terms of self-reinforcing dynamics~\cite{merton1968matthew,petersen2014reputation,bol2018matthew}, suggesting that an author's early success or fortune translate -- through a process of cumulative advantage -- into higher chances of further future success. There are several notable manifestations of the Matthew effect in academia. For instance, faculty at US universities are significantly more likely to have at least one parent with a PhD compared to the general population~\cite{morgan2022socioeconomic}, and are very likely to have been trained in a small group of elite institutions~\cite{clauset2015systematic,wapman2022quantifying}. In a similar fashion, an author's career impact is often significantly correlated with the visibility and prestige of their early career mentors~\cite{ma2020mentorship} and/or coauthors~\cite{li2019early}, and the likelihood of publishing as senior author in top interdisciplinary venues often boils down to being `chaperoned' to such venues by well established scientists~\cite{sekara2018chaperone}. Quite naturally, this has resulted in increasing inequality, with already highly cited authors receiving a rising share of citations~\cite{nielsen2021global}, leading to the formation of `rich clubs'~\cite{li2019reciprocity}. These dynamics are responsible for an effective narrowing of the literature, i.e., a small minority of papers capturing most of the attention~\cite{evans2008electronic,varga2022narrowing}, and for a more unequal allocation of funding across research institutions~\cite{ma2015anatomy,aagaard2020concentration}. 

The above results would suggest that a researcher's future impact overly depends on the initial conditions of their career. This, in turn, would imply that mobility within the impact ranking of a given discipline is highly restricted, which may stifle scientific innovation and fairness.

On the other hand, a few studies have illustrated somewhat opposite mechanisms and trends, showing that sustained career impact can emerge from early career failures and challenges (e.g., near-miss grant applications~\cite{wang2019early} or working on an interdisciplinary subject with low recognition~\cite{sun2021interdisciplinary}), which would facilitate mobility in impact rankings.

Motivated by the above findings, in this paper we examine the temporal evolution of academic impact rankings, that is the ranking of authors in a given discipline based on the citations they accrue over a period of time -- we are fully aware that citations are not a comprehensive measure of academic impact, yet citation-based indicators are the standard in the literature as they are easily quantifiable~\cite{garfield1955citation,vanclay2012impact,fortunato2018science}. Our analysis will focus on \emph{cohorts} of authors whose careers began in the same year. Each author will be characterized by their position in the rank of their cohort, which we will monitor over the first ten career years. This allows us to compare authors that began their careers under similar `environmental' conditions (in terms, e.g., of funding availability, volume of publications, maturity of their discipline, etc.) and competing for the same attention pool (in terms of potential readership of their work), therefore discounting the heterogeneity that would arise from the simultaneous presence of authors from different `generations' and focusing only on within-cohort inequality and ranking mobility.

In the following, we examine the ranking mobility of authors in different strata of their cohort, and we find that mobility is the lowest for the top and bottom-ranked authors. We study the evolution of ranking mobility by implementing a simple model akin to a random walk, which allows us to quantify mobility in different disciplines and epochs in terms of the model's `diffusion' coefficient. We observe that over time author cohorts have experienced -- on average -- increasing mobility and decreasing inequality in their discipline's impact rankings. 

\section*{Results}
\subsection*{Exploring authors' impact ranking mobility}
We begin our analysis by quantifying an author's mobility in the scientific impact ranking of their discipline. To this end, we create author publication profiles by employing a high-precision name disambiguation algorithm on Web of Science data (WoS, Materials and Methods), yielding a total of $5,194,173$ authors active in $57$ different disciplines spanning four macro areas of WoS data, i.e., Life Sciences \& Medicine, Physical Sciences, Social Sciences, and Technology (SI Appendix, Table.~\ref{tab:ScoreRules}). In line with numerous studies on scientific careers (e.g.,~\cite{sinatra2016quantifying}), we quantify the impact of each publication with the number of citations received within $5$ years after publication ($c_5$). We quantify an author's aggregate impact over a period of time as the sum of the $c_5$ scores of their publications during such period.

\begin{figure*}[ht!]
\centering
    \includegraphics[width=18cm]{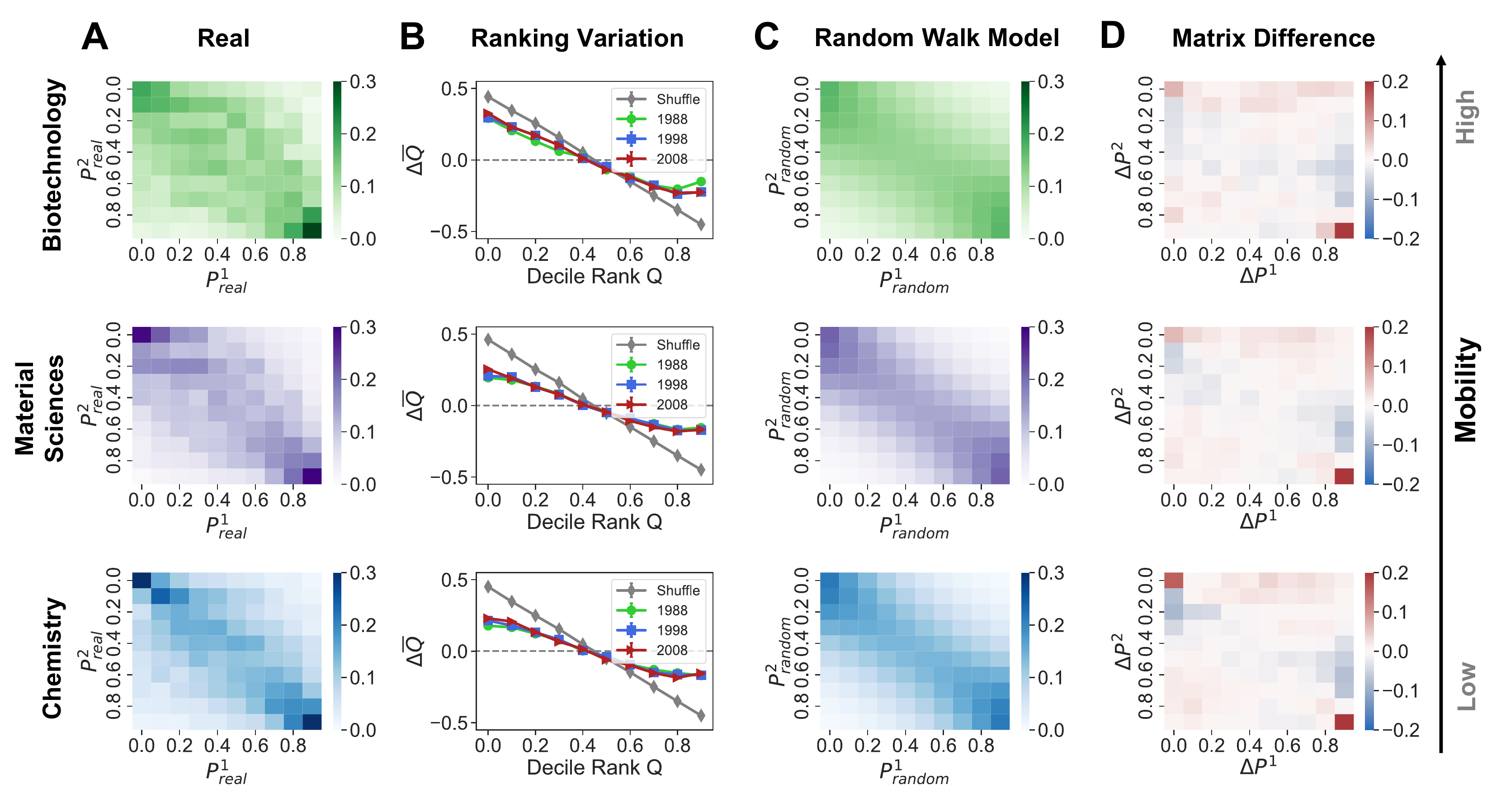}
    \caption{Author impact ranking mobility in Biotechnology, Materials Sciences, and Chemistry. (A) Transition matrix of author impact rankings between the first and second five years of their careers. Here, we consider the cohorts of authors who started their careers in the year $2000$, i.e., with first and second five career years covering 2000-2004 and 2005-2009, respectively. The scientific impact of an author over a period of time is calculated as the total number of citations received (within five years of publication) by their papers published over that period. Based on that we compute the decile rank of an author in their discipline during the first and second five career years, which we indicate with $P_{real}^1$ and $P_{real}^2$. (B) Average variation of impact ranking mobility $\Delta Q$ for each author percentile rank group $Q$. Results are shown for three different cohorts of authors, starting their careers in $1988$, $1998$, and $2008$, respectively. These are compared with the results from a null model obtained by randomly reshuffling the impact of authors during their second five career years. Error bars represent the standard error of the mean. (C) Transition probability matrix of authors in our random walk model (see~Eq.(\ref{eq:rw})). The optimal values of $D$ for Biotechnology, Materials Sciences, and Chemistry are equal to $0.35$, $0.22$, and $0.19$ respectively, implying that the mobility of authors in these disciplines decreases from top to bottom in the figure. (D) Differences between the empirical transition matrices (column A) and those obtained from the random walk model (column C).}
    \label{fig:AuthorMobility}
\end{figure*}

We group the authors in our dataset into cohorts based on their career starting year, which we identify with the year of their first publication on record, and examine their impact mobility between the first five years and second five years (i.e., years six to ten) of their careers. Namely, we rank the authors in each cohort based on their aggregate impact over their first five and second five career years, and divide both rankings into deciles. Note that only authors who have published at least one paper in each five year window are considered in our analysis. We then characterise impact mobility between the two time windows by computing $10 \times 10$ column-stochastic transition matrices, with entries given by the empirically estimated probability of an author moving from one decile to another. Fig.~\ref{fig:AuthorMobility}A shows the heat-maps of the transition matrices for author cohorts (with careers staring in $2000$) in three disciplines, i.e., Biotechnology, Materials Sciences and Chemistry. Consistently with those examples, most disciplines are characterised by transition matrices displaying concentration around the diagonal, indicating that most authors typically remain in the same decile or move to an adjacent one. Notably, in most disciplines we observe comparably large probabilities at each end of the diagonal in transition matrices, signalling a particularly high stability of the top and bottom of the impact ranking over time.

\begin{figure*}[ht!]
\centering
    \includegraphics[width=17cm]{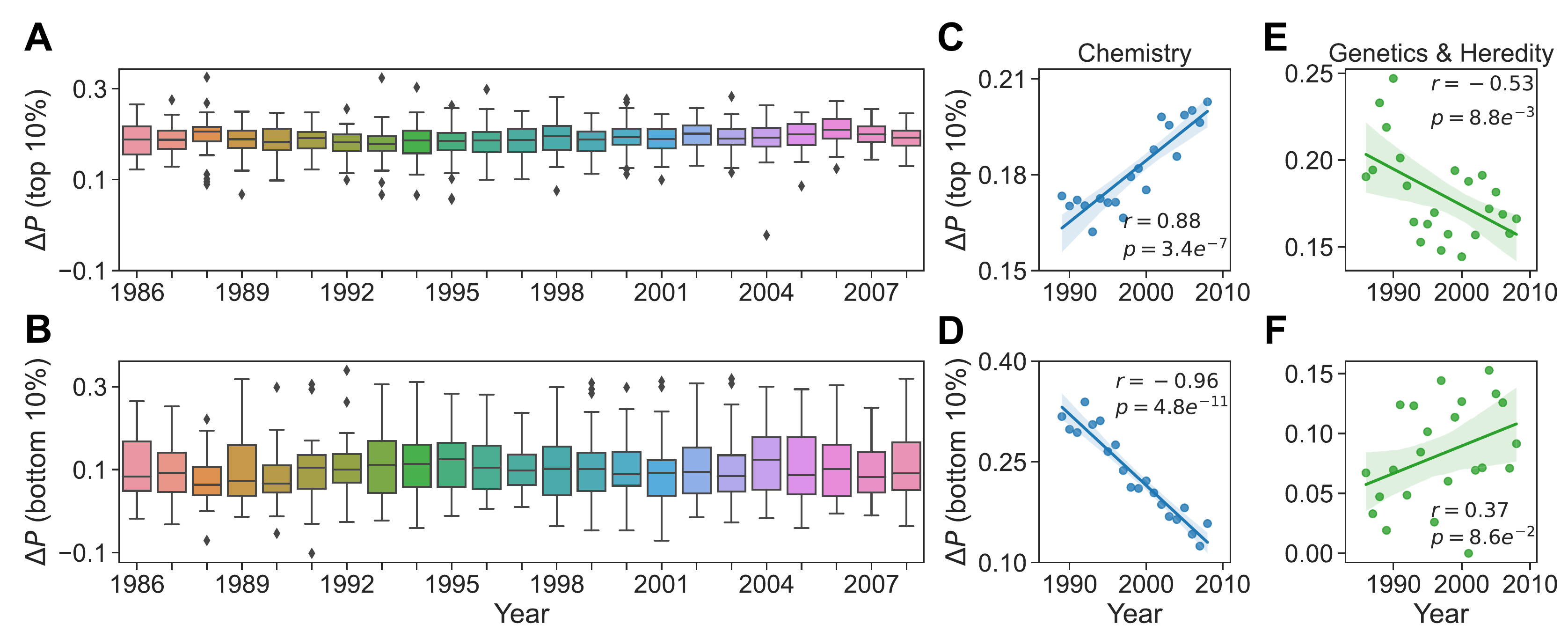}
    \caption{Evolution of mobility for authors in the top and bottom $10\%$ of impact rankings of their discipline. We show box plots of the overall evolution of $\Delta P$ for authors in the (A) top $10\%$ and (B) bottom $10\%$ of impact rankings obtained by aggregating over all disciplines. $\Delta P$ measures the difference of probability between real transition matrices and those obtained from the random walk model (Eq.(\ref{eq:rw})). We then report examples of different temporal evolutions of $\Delta P$ for different disciplines. Compared to the random walk model, the top $10\%$ of the impact ranking in Chemistry has become more stable over time (C), whereas the bottom $10\%$ has become more mobile (D). The opposite holds in the case of Genetics and Heredity, where stability has decreased in the top $10\%$ (E) and increased in the bottom $10\%$ (F). In figures (C)-(F) the solid lines and shaded areas represent regression lines and $95\%$ confidence level intervals, respectively. The Pearson coefficients -- and the corresponding $p$-values -- obtained from the regressions are shown in the figures.}
    \label{fig:Evolution_TopBottom} 
\end{figure*} 

To check whether these phenomena are common among authors entering the academic workforce at different times, we compute the average impact ranking mobility for authors starting their careers in $1988$, $1998$, and $2008$, respectively. We quantify impact mobility as the difference $\Delta Q$ between an author's decile rank in the second and first five career years, with $\Delta Q > 0$ ($\Delta Q < 0$) indicating that an author's ranking has improved (dropped) in the second time window. We compare these results with a null model obtained by randomly reshuffling the authors' impact in the second five career years, so as to completely decorrelate it with respect to impact over the first five years. Fig.~\ref{fig:AuthorMobility}B shows the average impact ranking mobility $\Delta \overline{Q}$ as a function of an author's decile rank $Q$ in their first five career years, as obtained both from the data and from the null model. As one would intuitively expect, on average authors located at the bottom of the ranking tend to move up ($\Delta \overline{Q} > 0$), whereas those at the top tend to move down ($\Delta \overline{Q} < 0$). However, the magnitude of such movements is remarkably different from the null model baseline. More precisely, we observe that the null model leads to an apparent linear relationship between $\Delta \overline{Q}$ and $Q$ with a slope roughly equal to $1$. Conversely, the corresponding relationship in the data is much more flat, especially for those at the top and bottom of the ranking, further highlighting the stability of such portions in impact rankings.

\begin{figure*}[ht!]
\centering
    \includegraphics[width=16cm]{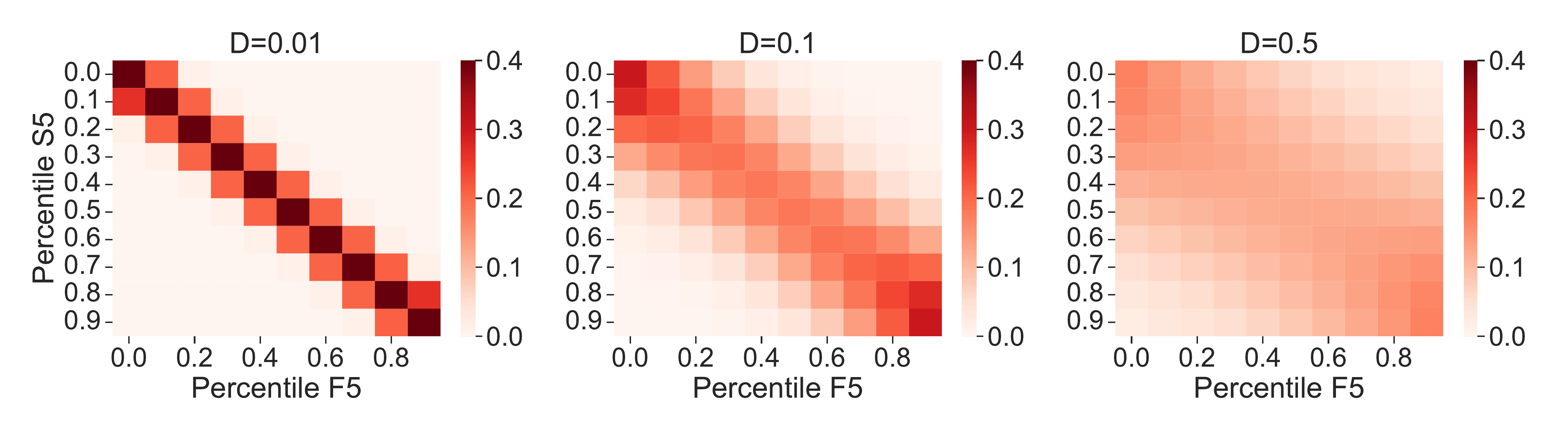}
    \caption{Transition matrices obtained from the random walk model for different values of the diffusion parameter $D$. As $D$ increases, the transition probabilities become more uniform, indicating an overall higher mobility of authors in impact rankings.}
    \label{fig:Theoretical_Mobility}
\end{figure*}

To further characterise author impact ranking mobility, we develop a simple model akin to a random walk aimed at capturing the aforementioned observed tendency of most authors to remain in the same decile or move to an adjacent one. Specifically, we assume the probability of an author moving from decile $j$ to decile $i$ to be given by
\begin{equation} \label{eq:rw}
P_{ij} = e^{-\Delta_{ij}^2/D} / \sum_\ell e^{-\Delta_{\ell j}^2/D}
\end{equation}
where $\Delta_{ij}$ is the distance between the two deciles, with $\Delta_{ii} = 0, \ \forall i$. In the random walk analogy, the parameter $D$ in~Eq.(\ref{eq:rw}) plays the role of the diffusion coefficient, controlling to which degree authors are able to ``diffuse'' towards higher/lower deciles of impact rankings. The higher the value of $D$, the more uniform the probabilities in a discipline's transition matrix, resulting in higher mobility (Fig.~\ref{fig:Theoretical_Mobility}). With a given value of $D$, we are able to compute all the entries in transition matrix successively. To capture and compare the overall mobility of author cohorts, we fit the optimal value of $D$ for authors in each discipline and year by minimising the Frobenius norm of the difference between the transition matrices of the real data and those of the random walk model.

\begin{figure*}[ht!]
\centering
    \includegraphics[width=18cm]{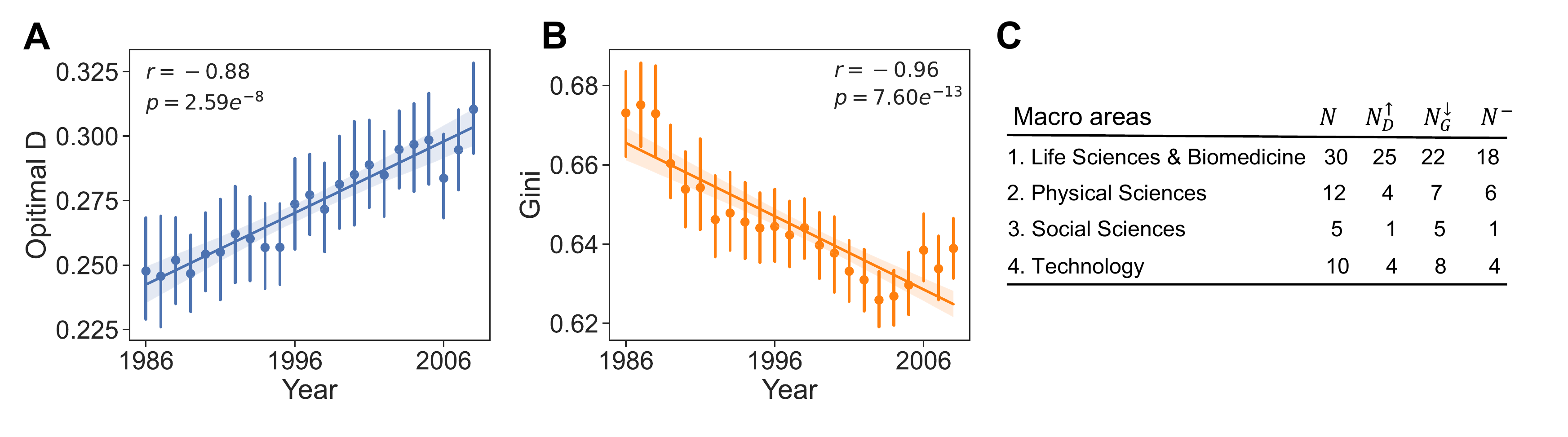}
    \caption{Evolution of mobility in impact rankings and inequality in impact distributions. (A) Mobility of author impact rankings (measured in terms of optimal $D$, see~Eq.(\ref{eq:rw})) for cohorts of authors starting their careers between $1986$ and $2008$. (B) Gini coefficients of impact distributions for cohorts of authors starting their careers between $1986$ and $2008$. The Gini coefficient is calculated based on the cumulative number of citations authors have received from the papers published during their first five career years, within five years of publication. In subplots A and B, the solid line and the shaded area represent regression lines and $95\%$ confidence level intervals, respectively. The error bars denote the $95\%$ confidence intervals of the distribution of optimal $D$. Each regression has also been annotated with the corresponding Pearson correlation coefficient $r$ and its $p$-value. (C) Number of disciplines with statistically significant trends over time within each macro area (with Pearson correlation coefficient p value < 0.1). $N$ denotes the number of disciplines in each macro area. $N_{D}^{\uparrow}$ and $N_{G}^{\downarrow}$ respectively denote the number of disciplines with a statistically significant increasing trend in mobility and a decreasing trend in inequality. $N^{-}$ counts the number of disciplines with a statistically significant negative correlation between mobility and inequality.}
    \label{fig:Evolution_Gini_Mobility}
\end{figure*} 

In Fig.~\ref{fig:AuthorMobility}C we show the transition matrices for Biotechnology, Materials Sciences, and Chemistry at the optimal values of $D$, equal to $0.35$, $0.22$, and $0.19$, respectively. From top to bottom, we can observe that -- as expected based on such values of $D$ -- transition matrices become less uniform, with more probability concentrated around the diagonal, signalling a lower mobility. Fig.~\ref{fig:AuthorMobility}D depicts the element-wise differences $\Delta P$ between empirical transition matrices and those obtained from optimal values of $D$. Generally speaking, we find that the random walk model captures well the overall characteristics of author ranking mobility (with most values of $\Delta P$ nearly equal to $0$). In line with the results shown in~Fig.~\ref{fig:AuthorMobility}B, the model underestimates the probabilities for authors in top and bottom deciles to remain in such groups. 

\begin{figure*}[ht!]
\centering
    \includegraphics[width=18cm]{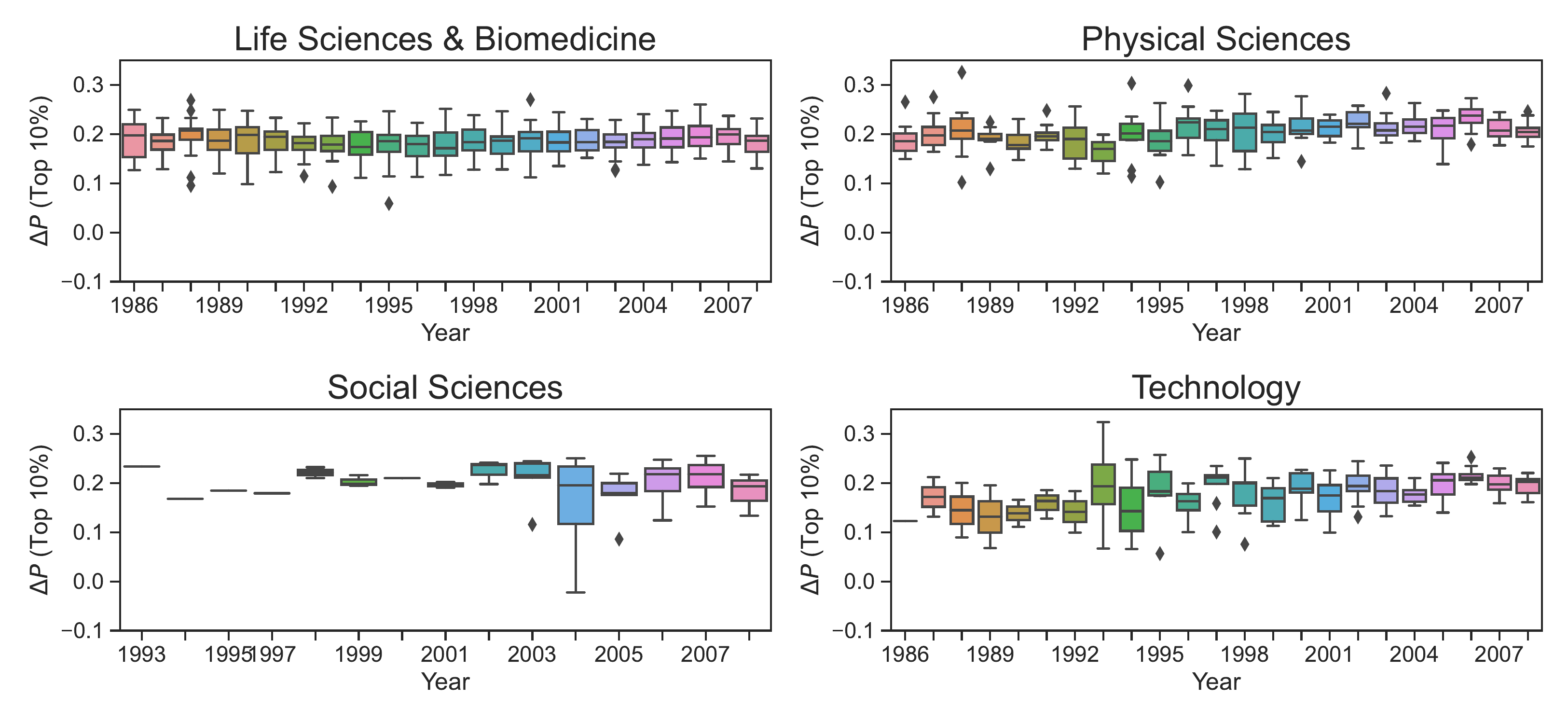}
    \caption{Box plots of the evolution of $\Delta P$ for authors ranked in the top $10\%$ of scientific impact in the four macro areas. $\Delta P$ is the difference of the transition probabilities between the empirical transition matrix and the one obtained from the random walk model. We can see that for each of the macro areas, the average value of $\Delta P$ remains positive and remains almost constant during the whole time period. We also find that the average value of $\Delta P$ in the Physical Sciences ($\Delta \overline{P}=0.205$) across time is significantly higher than that in the Life Science \& Biomedicine ($\Delta \overline{P}=0.185$) and Technology ($\Delta \overline{P}=0.182$), suggesting that top-ranked authors in the Physical Sciences are less likely to change their impact ranking than those in Life Science \& Biomedicine and Technology. The significance level refers to two-tailed $t$-tests with $p<3.06e^{-8}$.}
    \label{fig:Top10_diff}
\end{figure*}

\begin{figure*}[ht!]
\centering
    \includegraphics[width=18cm]{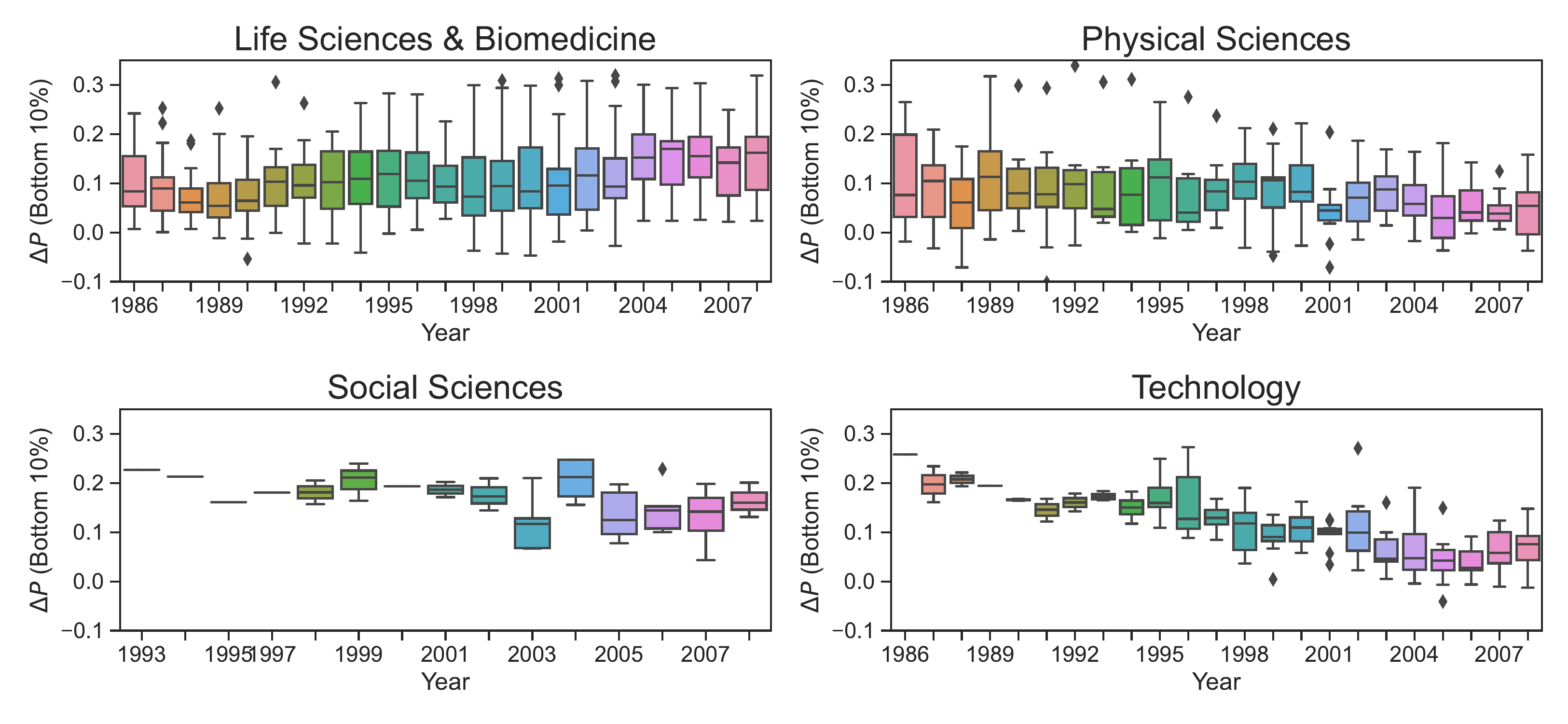}
    \caption{Box plots of the evolution of $\Delta P$ for authors ranked in the bottom $10\%$ of scientific impact in the four macro areas. $\Delta P$ is the difference of the transition probabilities between the empirical transition matrix and the one obtained from the random walk model. One can see that, in general, the average $\Delta P$ remains roughly constant across the whole period, except for Technology. The average values of $\Delta P$ for Life Sciences \& Biomedicine, Physical Sciences, Social Sciences and Technology are $0.12$, $0.08$, $0.15$ and $0.10$, respectively.}
    \label{fig:Bottom10_diff}
\end{figure*}

We then proceed to test whether the excess of stability in the top and bottom $10\%$ is persistent over time. In order to do that, we report the difference between transition probabilities $\Delta P$ in real data and in the random walk model across all disciplines at the aggregate level in Fig.~\ref{fig:Evolution_TopBottom}A-B. Overall, the average values of $\Delta P$ remain positive and roughly constant throughout our entire period of observation, indicating that the stability at the top and bottom of impact rankings has been consistently higher, and that the gap has not increased over time. However, we find a higher degree of stability at the top of impact rankings, as testified by a narrower distribution of $\Delta P$ with a higher average value ($0.19$) with respect to the bottom of impact rankings (average $\Delta P$ equal to $0.10$). We further validate these results at the level of WoS macro areas, arriving at the same conclusions, but the extent of the gaps between the data and the random walk model are different between the macro research areas (see Fig.~\ref{fig:Top10_diff} and Fig.~\ref{fig:Bottom10_diff}). Notably, for authors in the top $10\%$, the average value of $\Delta P$ in Physical Sciences is statistically higher than that of the other three macro areas, while for the bottom $10\%$ authors, the average value of $\Delta P$ in Physical Sciences are the lowest ($p<0.001$, two tailed $t$-tests). This suggests that authors in Physical Sciences generally have a very stable top portion of the impact ranking, but a relatively flexible bottom portion compared with other macro areas.

\begin{figure*}[ht!]
\centering
    \includegraphics[width=16cm]{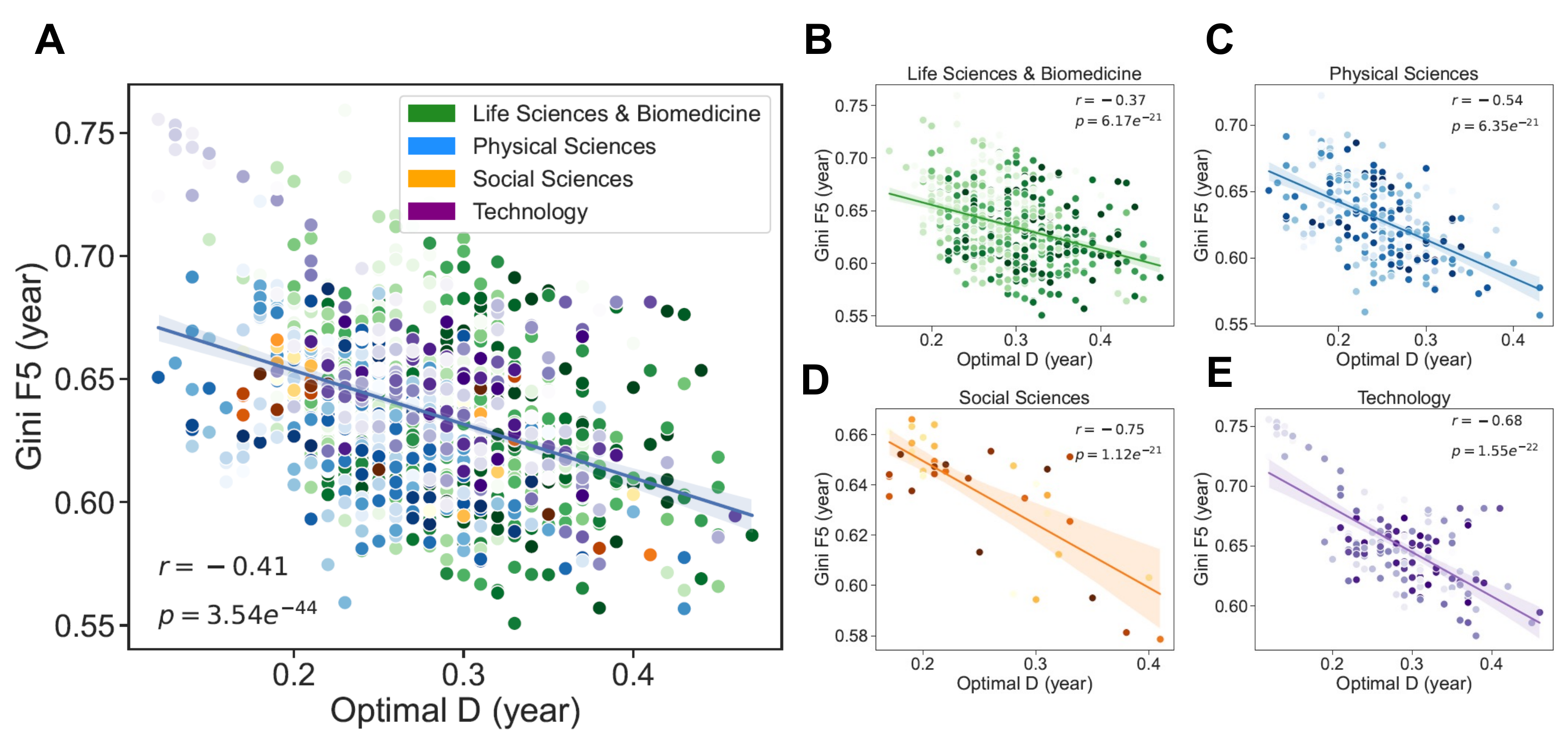}
    \caption{Correlation between mobility in impact rankings and inequality in impact distributions for all author cohorts across all disciplines (A), and for author cohorts in Life science and Biomedicine (B), Physics Sciences (C), Social Sciences (D) and Technology (E). The Gini coefficient is calculated based on the cumulative number of citations authors have received from the papers published during their first five career years, within five years of publication. In each subplot, each circle represents an author cohort in a given discipline. The shading of the circles, from light to dark, indicates the year in which a cohort of authors started their careers, from $1986$ to $2008$. The solid line and the shaded area indicate regression lines and $95\%$ confidence level interval, respectively. Each regression has also been annotated with the corresponding Pearson correlation coefficient $r$ and its $p$-value.}
    \label{fig:Correlation_Gini_Mobility} 
\end{figure*} 

\begin{figure*}[ht!]
\centering
    \includegraphics[width=16cm]{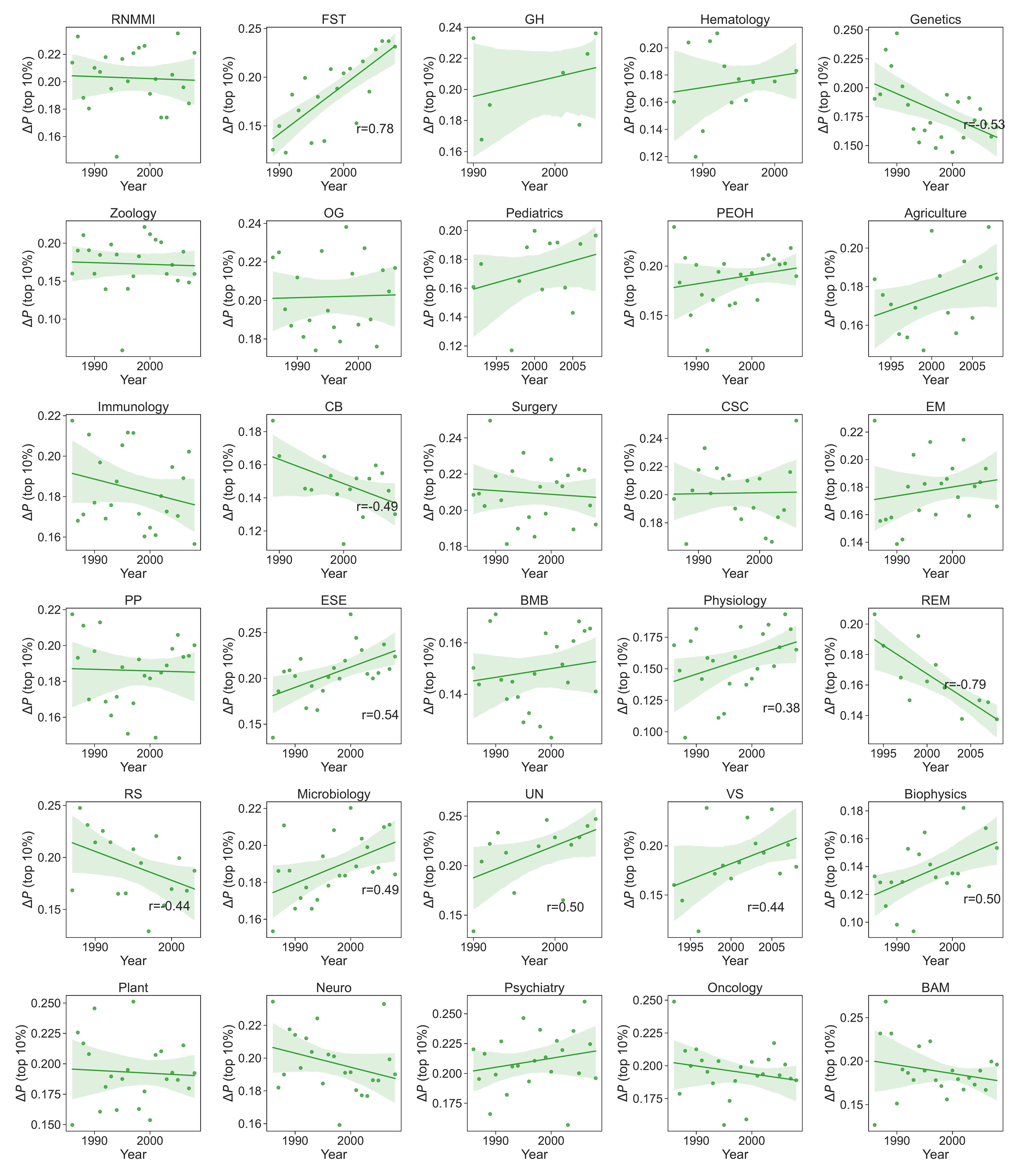}
    \caption{Evolution of the difference in transition probability for the top $10\%$ most cited authors in each discipline within Life Sciences \& Biomedicine. The solid line and the shaded area indicate the regression line and the $95\%$ confidence interval, respectively. The regressions with $p<0.1$ have been annotated with the corresponding Pearson’s $r$.}
    \label{fig:Top10_diff_life}
\end{figure*}

\begin{figure*}[ht!]
\centering
    \includegraphics[width=16cm]{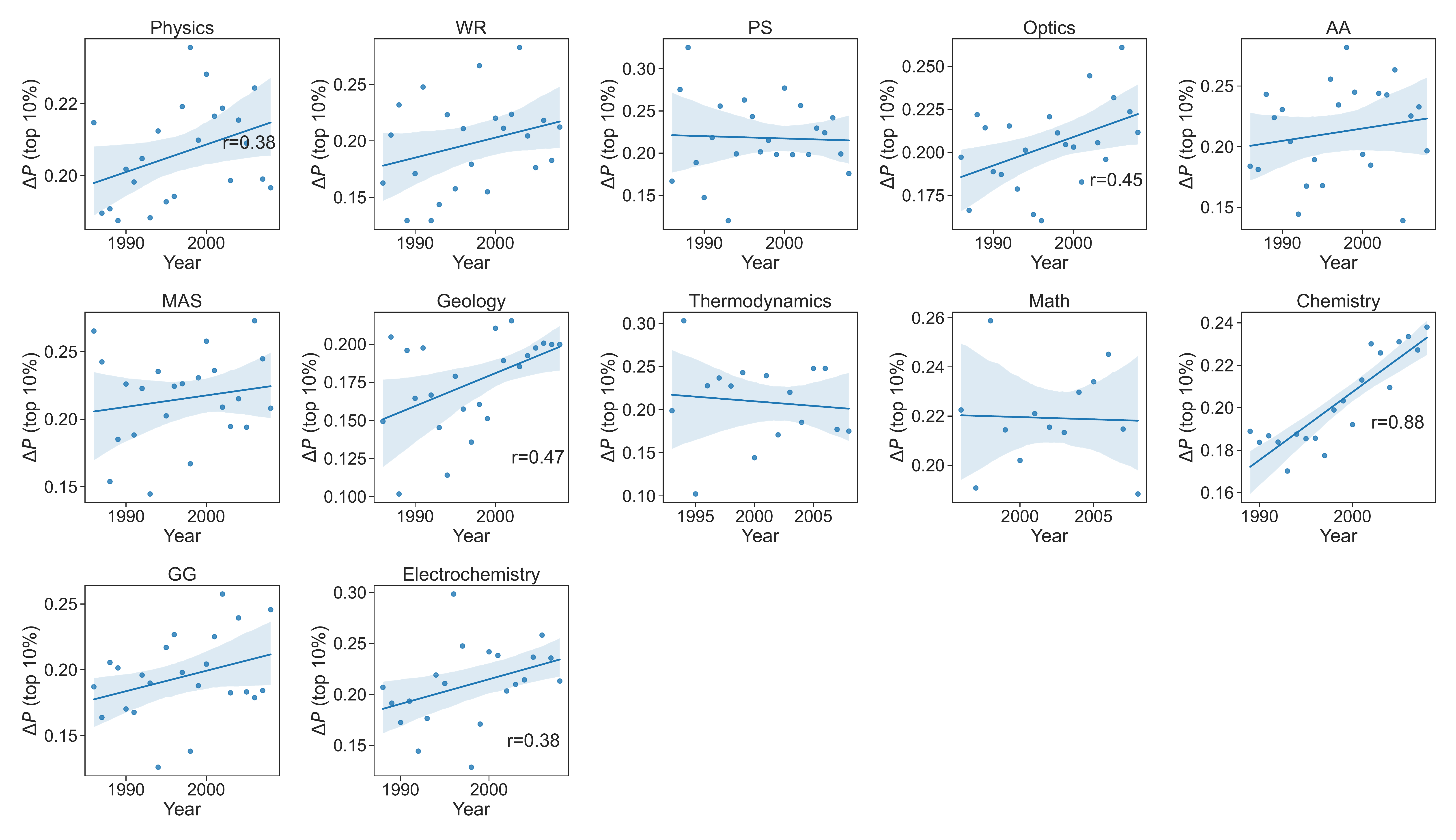}
    \caption{Evolution of the difference in transition probability for the top $10\%$ most cited authors in each discipline within the Physical Sciences. The solid line and the shaded area indicate the regression line and the $95\%$ confidence interval, respectively. The regressions with $p<0.1$ have been annotated with the corresponding Pearson’s $r$.}
    \label{fig:Top10_diff_physics}
\end{figure*}

\begin{figure*}[ht!]
\centering
    \includegraphics[width=16cm]{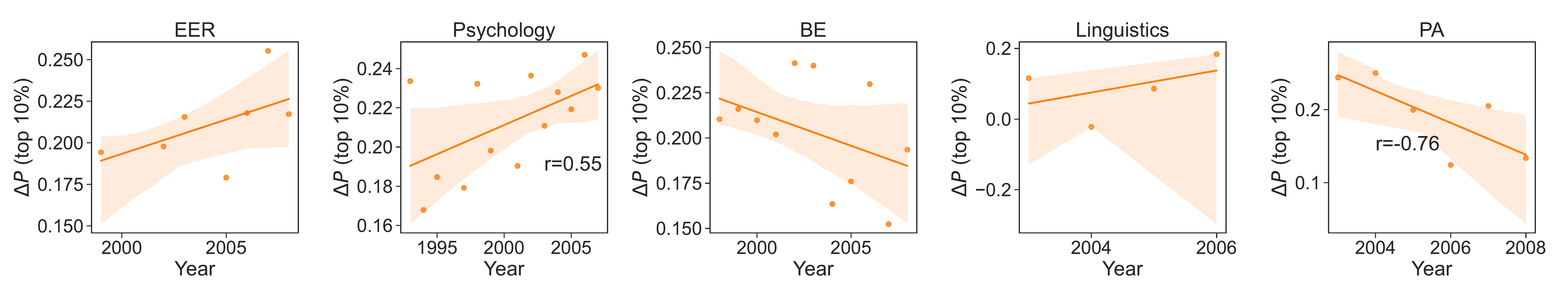}
    \caption{Evolution of the difference in transition probability for the top $10\%$ most cited authors in each discipline within the Social Sciences. The solid line and the shaded area indicate the regression line and the $95\%$ confidence interval, respectively. The regressions with $p<0.1$ have been annotated with the corresponding Pearson’s $r$.}
    \label{fig:Top10_diff_social}
\end{figure*}

\begin{figure*}[ht!]
\centering
    \includegraphics[width=16cm]{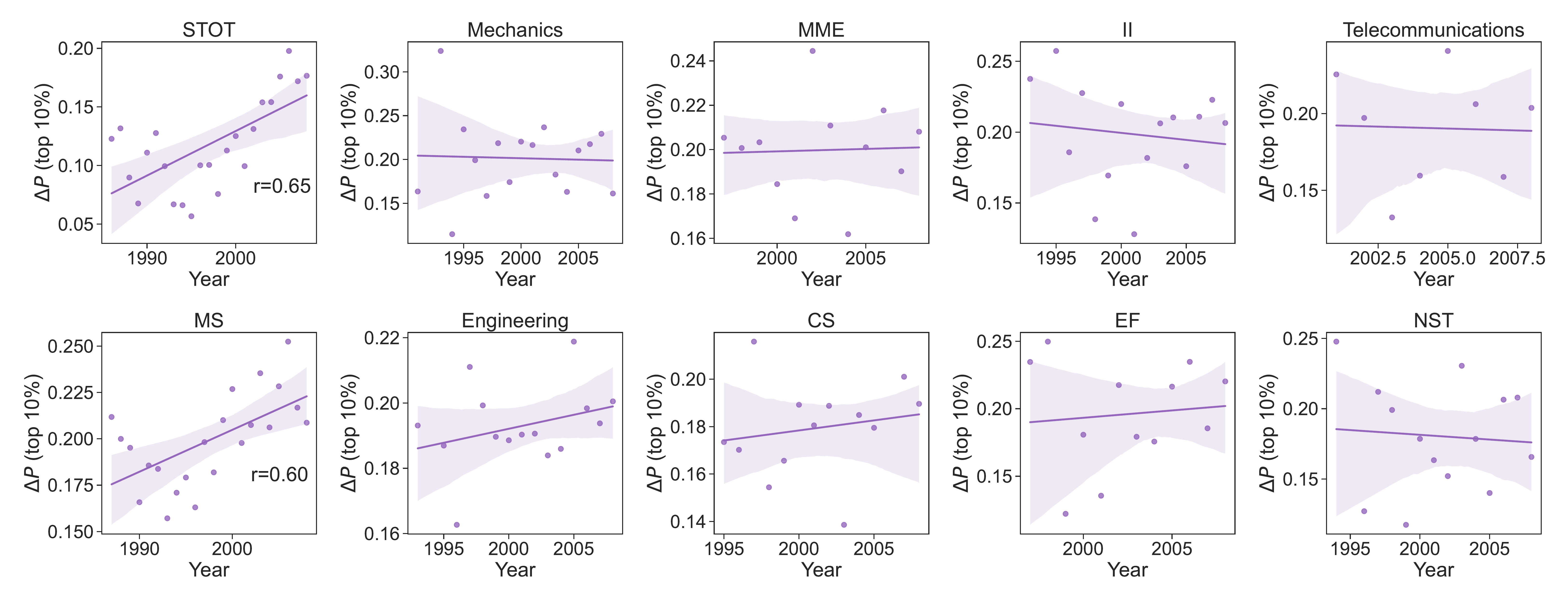}
    \caption{Evolution of the difference in transition probability for the top $10\%$ most cited authors in each discipline within Technology. The solid line and the shaded area indicate the regression line and the $95\%$ confidence interval, respectively. The regressions with $p<0.1$ have been annotated with the corresponding Pearson’s $r$.}
    \label{fig:Top10_diff_tech}
\end{figure*}

\begin{figure*}[ht!]
\centering
    \includegraphics[width=16cm]{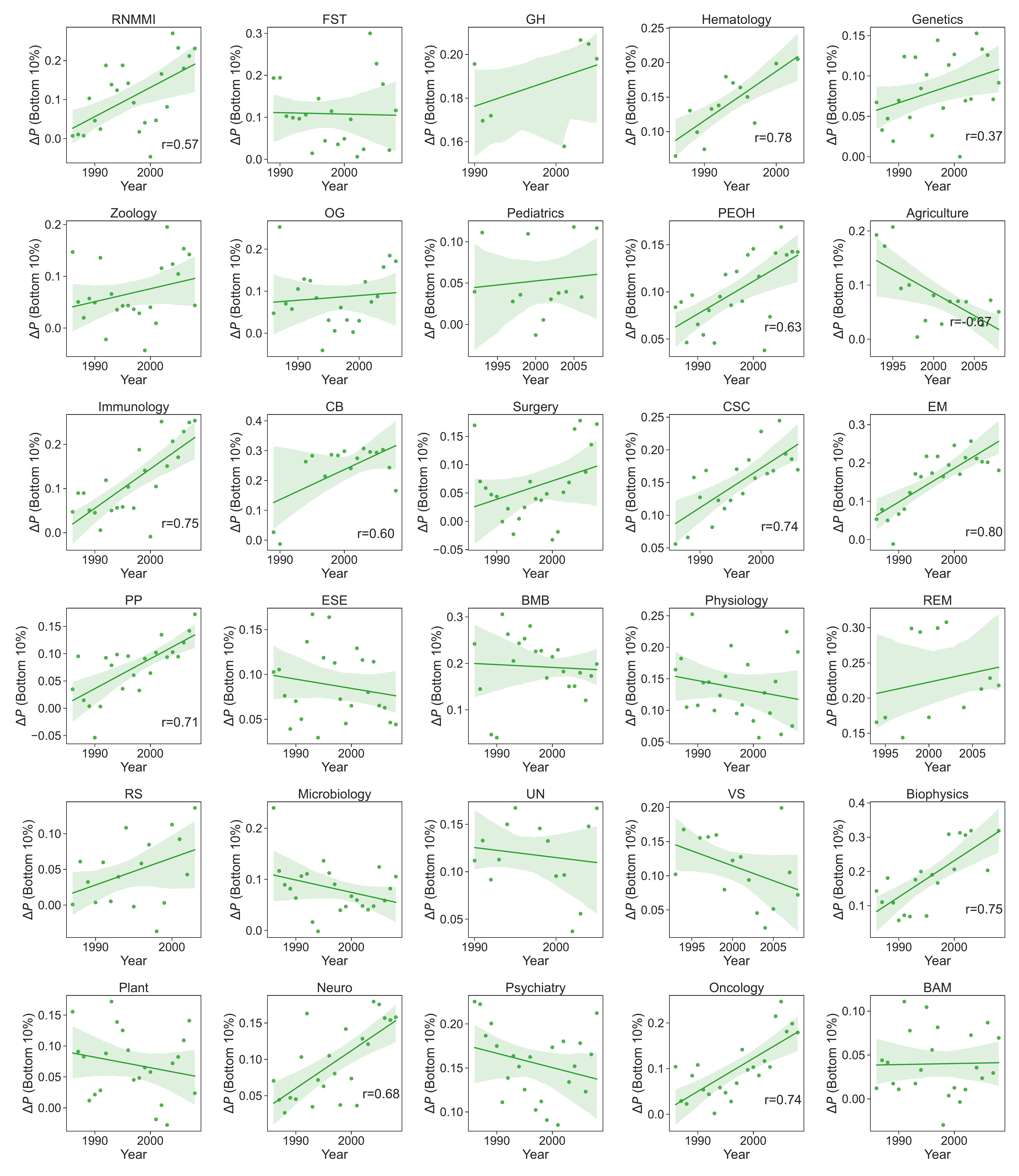}
    \caption{Evolution of the difference in transition probability for the $10\%$ least cited authors in Life Sciences \& Biomedicine. The solid line and the shaded area indicate the regression line and the $95\%$ confidence interval, respectively. The regressions with $p<0.1$ have been annotated with the corresponding Pearson’s $r$.}
    \label{fig:Bottom_10_diff_life}
\end{figure*}

\begin{figure*}[ht!]
\centering
    \includegraphics[width=16cm]{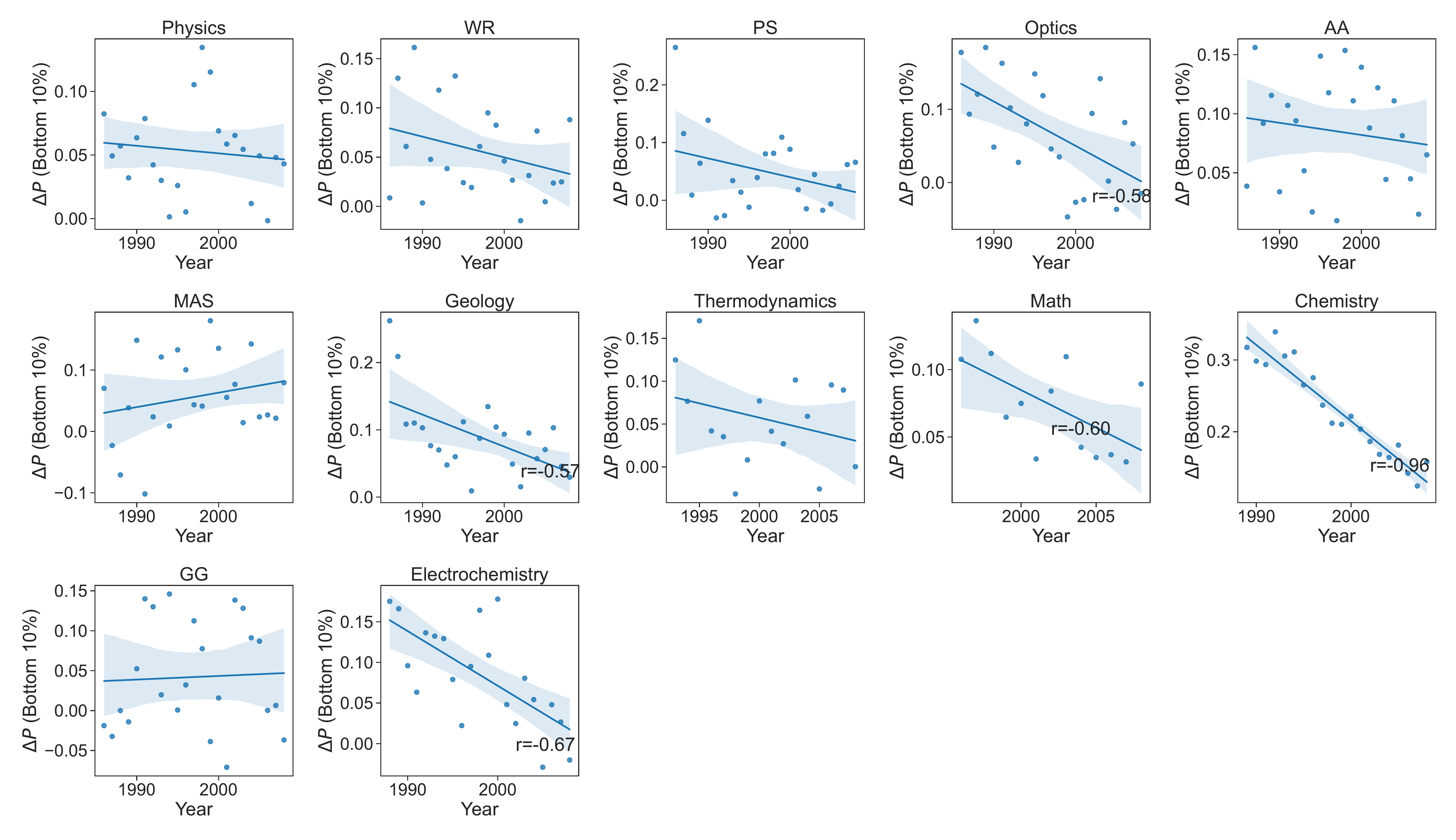}
    \caption{Evolution of the difference in transition probability for the $10\%$ least cited authors in the Physical Sciences. The solid line and the shaded area indicate the regression line and the $95\%$ confidence interval, respectively. The regressions with $p<0.1$ have been annotated with the corresponding Pearson’s $r$.}
    \label{fig:Bottom_10_diff_physics}
\end{figure*}

\begin{figure*}[ht!]
\centering
    \includegraphics[width=16cm]{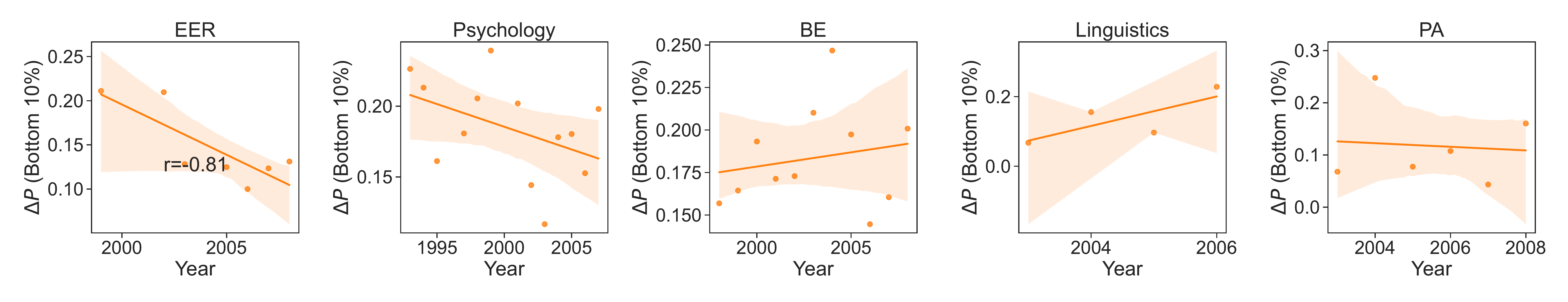}
    \caption{Evolution of the difference in transition probability for the $10\%$ least cited authors in the Social Sciences. The solid line and the shaded area indicate the regression line and the $95\%$ confidence interval, respectively. The regressions with $p<0.1$ have been annotated with the corresponding Pearson’s $r$.}
    \label{fig:Bottom_10_diff_social}
\end{figure*}

\begin{figure*}[ht!]
\centering
    \includegraphics[width=16cm]{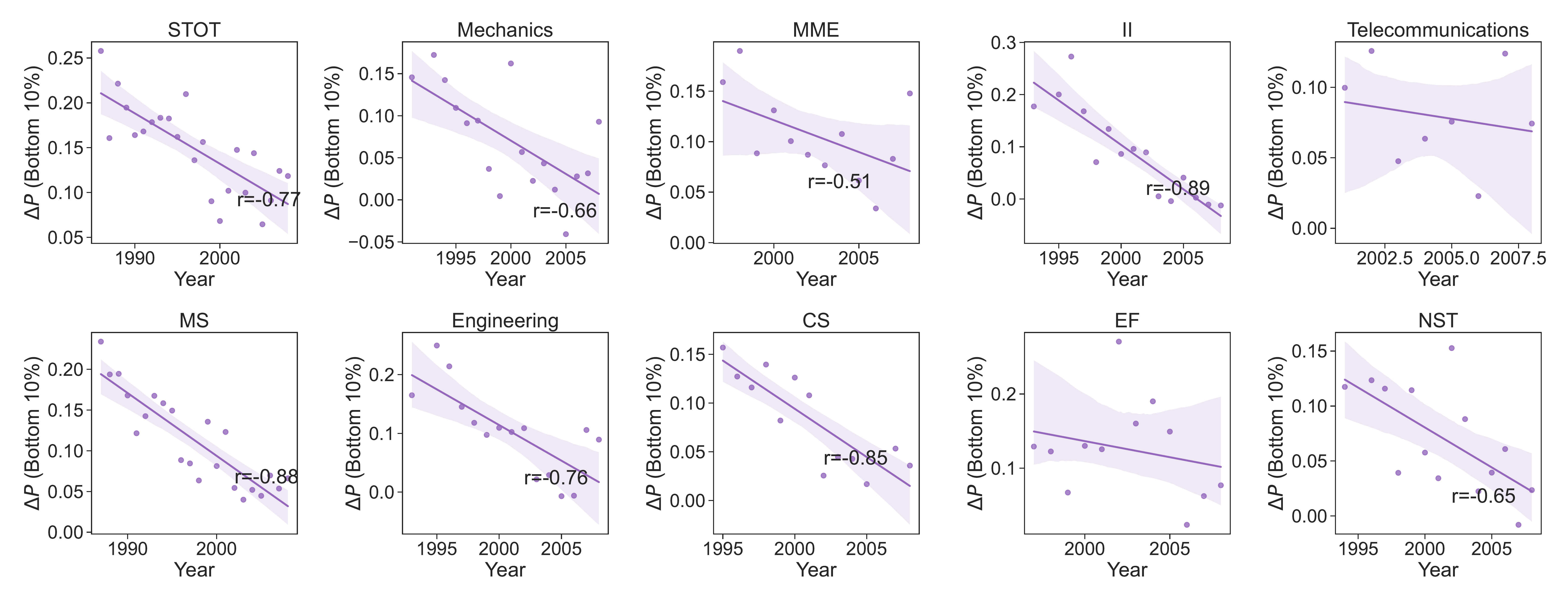}
    \caption{Evolution of the difference in transition probability for the $10\%$ least cited authors in Technology. The solid line and the shaded area indicate the regression line and the $95\%$ confidence interval, respectively. The regressions with $p<0.1$ have been annotated with the corresponding Pearson’s $r$.}
    \label{fig:Bottom_10_diff_tech}
\end{figure*}

Although the overall degree of differences remains roughly constant over time when aggregating over all disciplines or at the level of macro areas, we do observe some specific temporal trends at the level of individual disciplines (Fig.~\ref{fig:Top10_diff_life}-\ref{fig:Bottom_10_diff_tech}). For example, the gap $\Delta P$ for the top $10\%$ ranked authors in Chemistry has been steadily increasing from $1986$ to $2008$ (Fig.~\ref{fig:Evolution_TopBottom}C), while the gap for the bottom $10\%$ has been steadily decreasing (Fig.~\ref{fig:Evolution_TopBottom}D). Conversely, we observe opposite trends in Genetics and Heredity (Fig.~\ref{fig:Evolution_TopBottom} E-F), with a decline in stability for top-ranked authors, and an increase for bottom-ranked ones. Taken together, these findings demonstrate that -- on average -- top- and bottom-ranked authors consistently experience higher stability, but with notable differences in terms of temporal patterns across different disciplines.

\subsection*{Evolution of authors' mobility and inequality}
We now proceed to examine how mobility in author impact rankings has evolved over time and across disciplines. We estimate the mobility of different author cohorts by estimating the optimal value of the parameter $D$ in~Eq.(\ref{eq:rw}) for each discipline and year. As can be seen in Fig.~\ref{fig:Evolution_Gini_Mobility}A, there has been a significant increase in mobility over the two decades in our analysis. In other words, the later an author started their academic career the lower -- on average -- the likelihood that their future impact depends on their past impact. 

We hypothesize that higher (lower) levels of mobility in impact rankings should be accompanied by a lower (higher) level of concentration in the distribution of citations received by authors. To test this hypothesis, we first study the variation of impact concentration for author cohorts over time. We calculate the Gini coefficient (one of the most widely used indicators of inequality) for each author cohort and each discipline to gauge the concentration of author impact distributions. Here, an author's impact is defined as the total number of citations received for all the papers published in the first five years of their careers within 5 years of publication. Indeed, aggregating over all disciplines we observe that the overall concentration of impact at the level of cohorts has steadily decreased for authors who started their careers between 1986 and 2003, albeit with a rebound in impact inequality from 2003 onwards (Fig.~\ref{fig:Evolution_Gini_Mobility}B).

\begin{figure*}[ht!]
\centering
    \includegraphics[width=16cm]{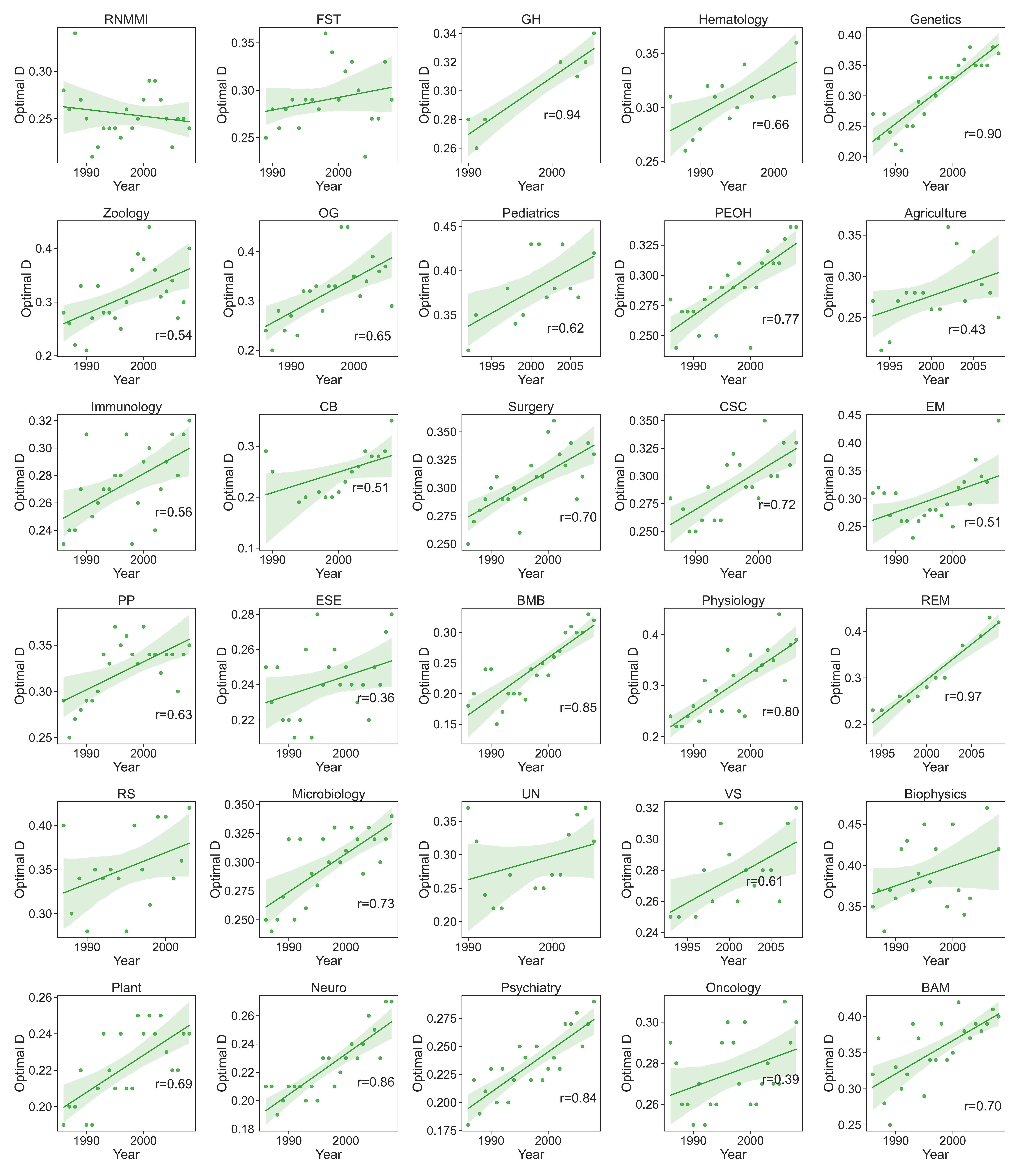}
    \caption{Author impact ranking mobility over time in the Life Sciences \& Biomedicine. The disciplines with significant increasing trends of mobility over time are marked with Pearson correlation coefficient $r$ ($p$-value < 0.1).}
    \label{fig:AuthorMobility_life}
\end{figure*}

\begin{figure*}[ht!]
\centering
    \includegraphics[width=16cm]{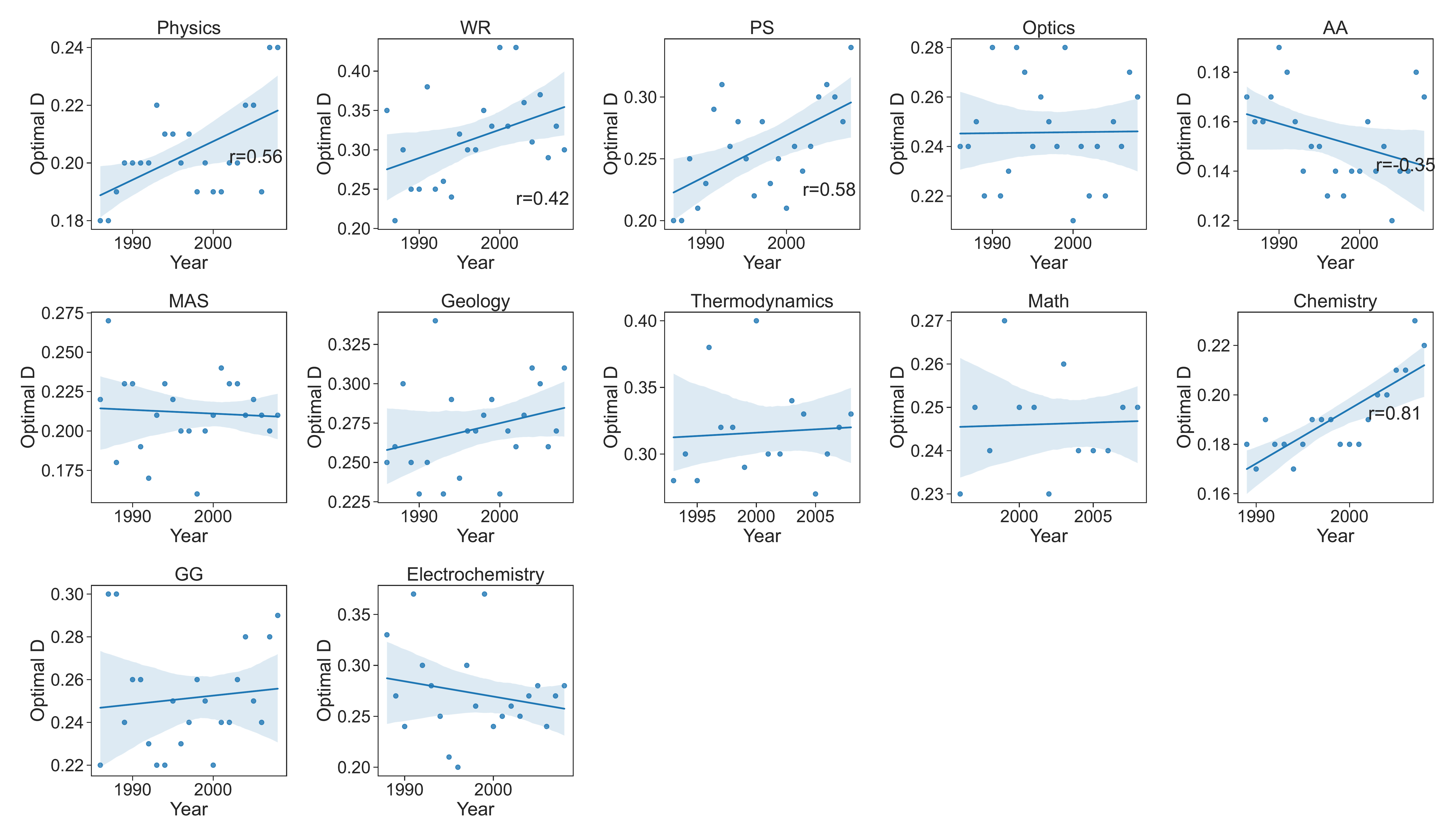}
    \caption{Author impact ranking mobility over time in the Physical Sciences. The disciplines with significant increasing trends of mobility over time are marked with Pearson correlation coefficient $r$ ($p$-value < 0.1).}
    \label{fig:AuthorMobility_physics}
\end{figure*}

\begin{figure*}[ht!]
\centering
    \includegraphics[width=16cm]{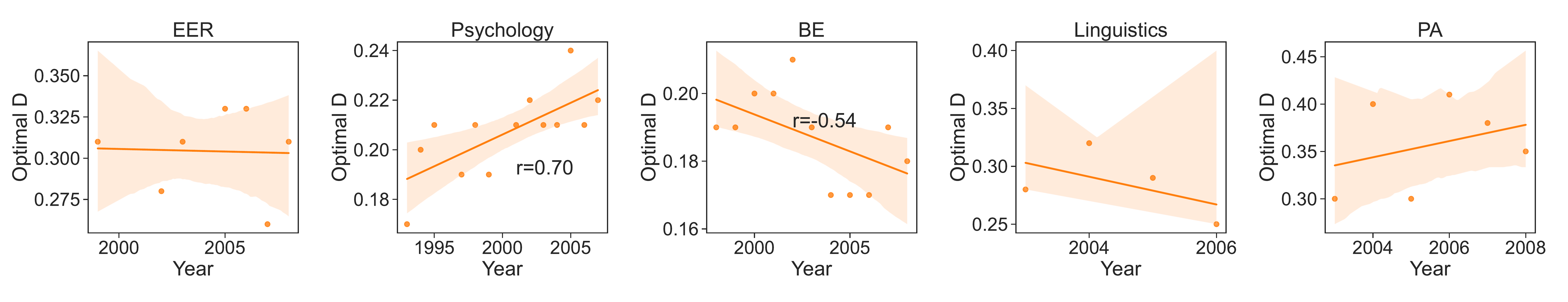}
    \caption{Author impact ranking mobility over time in the Social Sciences. The disciplines with significant increasing trends of mobility over time are marked with Pearson correlation coefficient $r$ ($p$-value < 0.1).}
    \label{fig:AuthorMobility_social}
\end{figure*}

\begin{figure*}[ht!]
\centering
    \includegraphics[width=16cm]{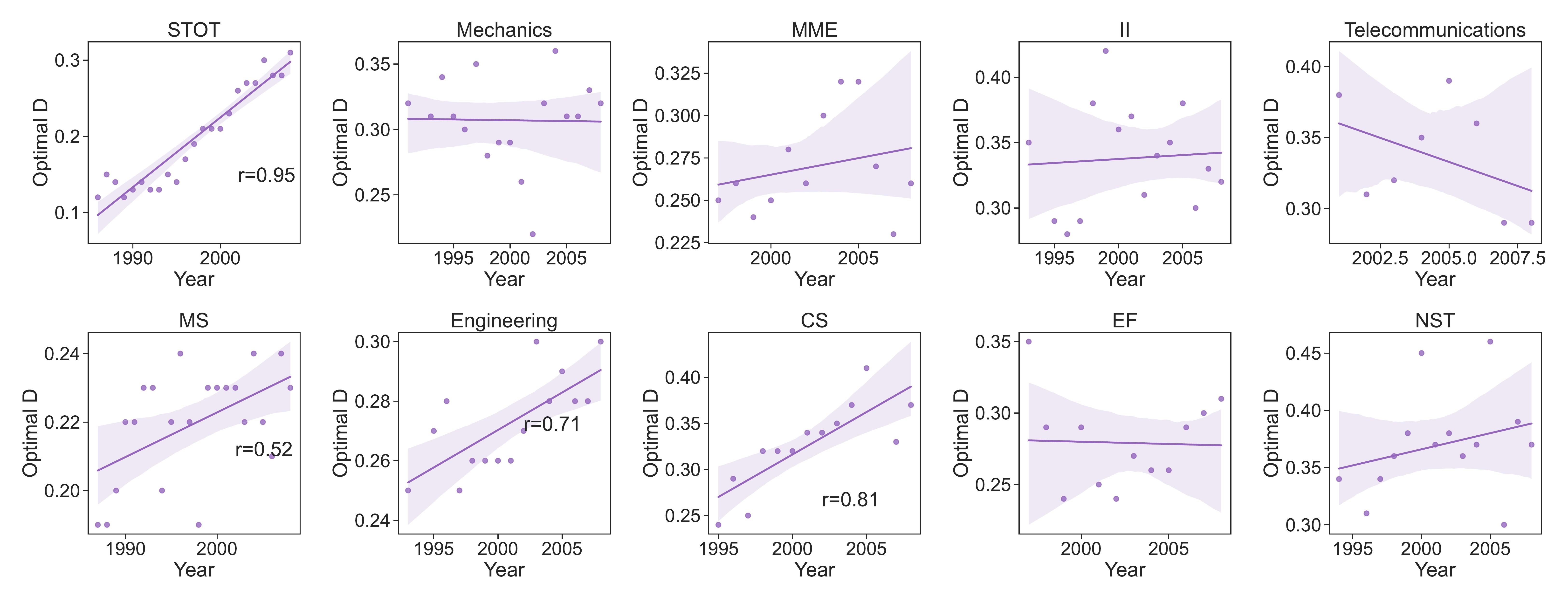}
    \caption{Author impact ranking mobility over time in Technology. The disciplines with significant increasing trends of mobility over time are marked with Pearson correlation coefficient $r$ ($p$-value < 0.1).}
    \label{fig:AuthorMobility_technology}
\end{figure*}

\begin{figure*}[ht!]
\centering
    \includegraphics[width=16cm]{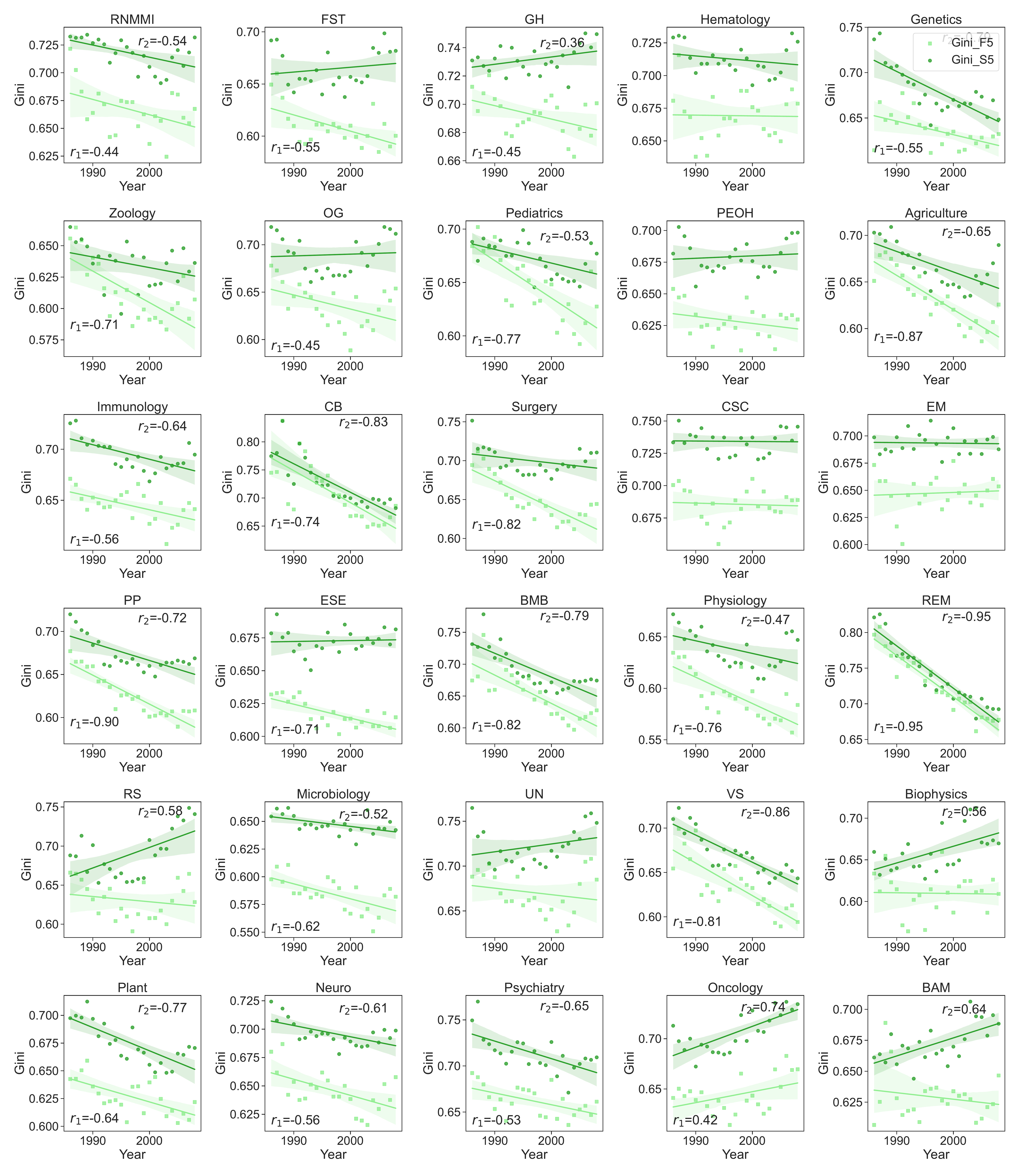}
    \caption{Gini coefficients of author citation distributions in the Life Sciences \& Biomedicine from $1986$ to $2008$. The Gini coefficients are calculated based on the distribution of the cumulative number of citations obtained by authors during the first five years (light green square) and the second five years (dark green circle) of their careers. The disciplines with significant increasing/decreasing trends of mobility over time (with $p$-value < 0.1) are marked with Pearson correlation coefficient $r_{1}$ (for first 5 years) and $r_{2}$ (for second 5 years).}
    \label{fig:AuthorGini_life}
\end{figure*}

\begin{figure*}[ht!]
\centering
    \includegraphics[width=16cm]{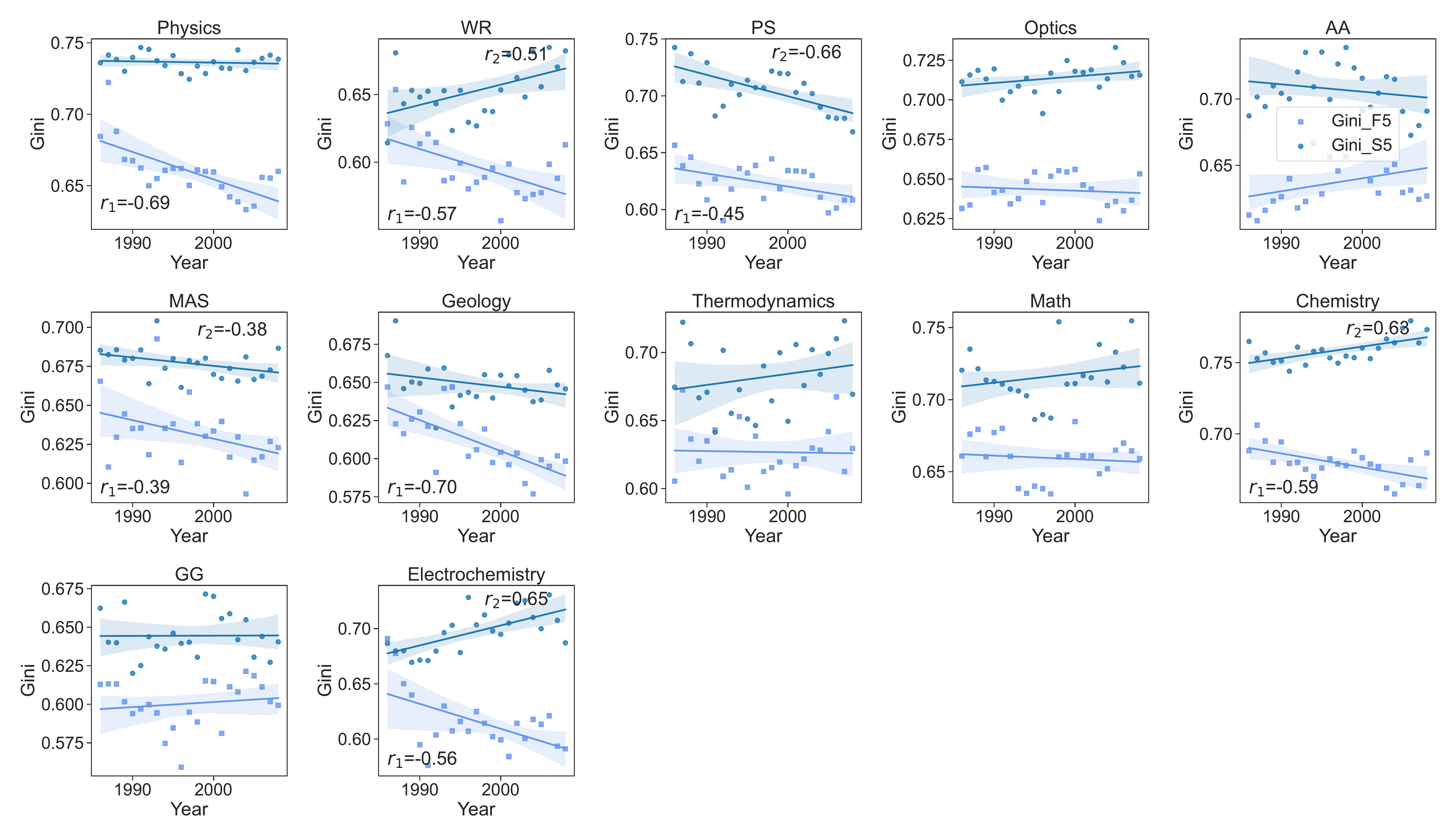}
    \caption{Gini coefficients of author citation distributions in the Physical Sciences from 1986 to 2008. The Gini coefficients are calculated based on the distribution of the cumulative number of citations obtained by authors during the first five years (light blue square) and the second five years (dark blue circle) of their careers. The disciplines with significant increasing/decreasing trends of mobility over time (with $p$-value < 0.1) are marked with Pearson correlation coefficient $r_{1}$ (for first 5 years) and $r_{2}$ (for second 5 years).}
    \label{fig:AuthorGini_physics}
\end{figure*}

\begin{figure*}[ht!]
\centering
    \includegraphics[width=16cm]{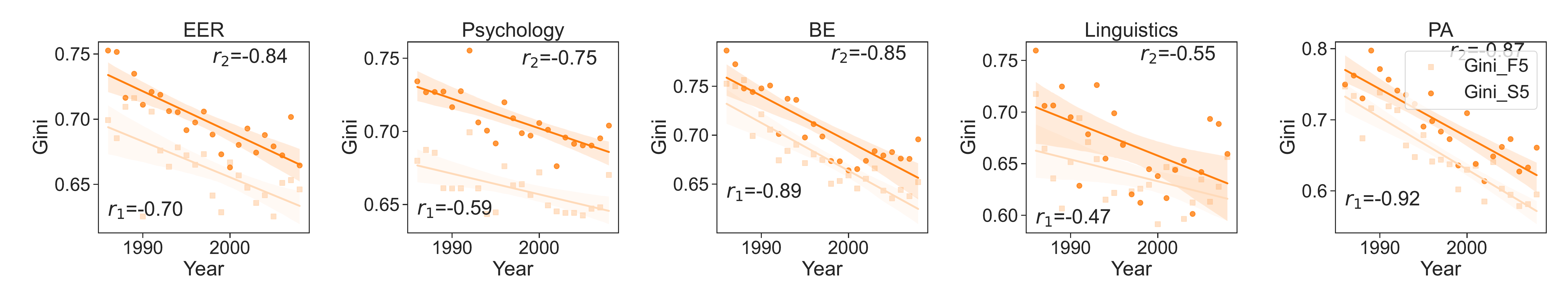}
    \caption{Gini coefficients of author citation distributions in the Social Sciences from 1986 to 2008. The Gini coefficients are calculated based on the distribution of the cumulative number of citations obtained by authors during the first five years (light orange square) and the second five years (dark orange circle) of their careers. The disciplines with significant increasing/decreasing trends of mobility over time (with $p$-value < 0.1) are marked with Pearson correlation coefficient $r_{1}$ (for first 5 years) and $r_{2}$ (for second 5 years).}
    \label{fig:AuthorGini_social}
\end{figure*}

\begin{figure*}[ht!]
\centering
    \includegraphics[width=16cm]{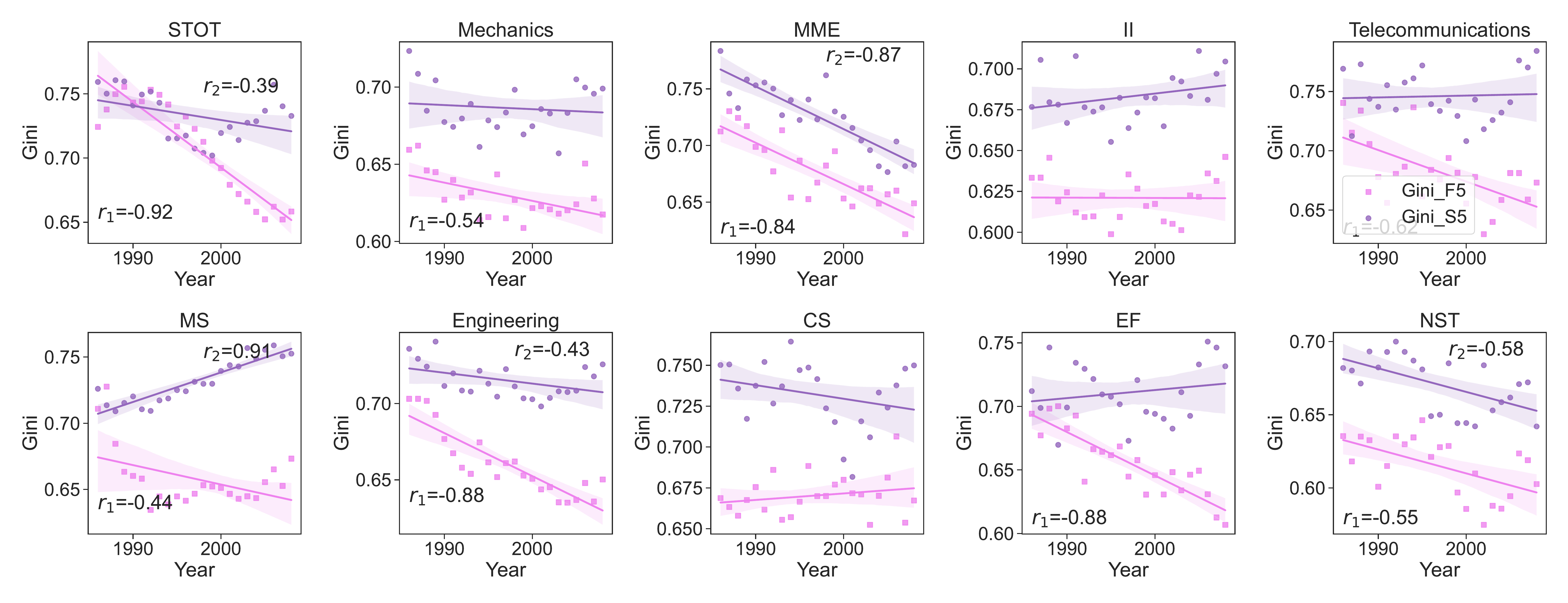}
    \caption{Gini coefficients of author citation distributions in Technology from 1986 to 2008.The Gini coefficients are calculated based on the distribution of the cumulative number of citations obtained by authors during the first five years (light pink square) and the second five years (dark purple circle) of their careers. The disciplines with significant increasing/decreasing trends of mobility over time (with $p$-value < 0.1) are marked with Pearson correlation coefficient $r_{1}$ (for first 5 years) and $r_{2}$ (for second 5 years).}
    \label{fig:AuthorGini_technology}
\end{figure*}

\begin{figure*}[ht!]
\centering
    \includegraphics[width=16cm]{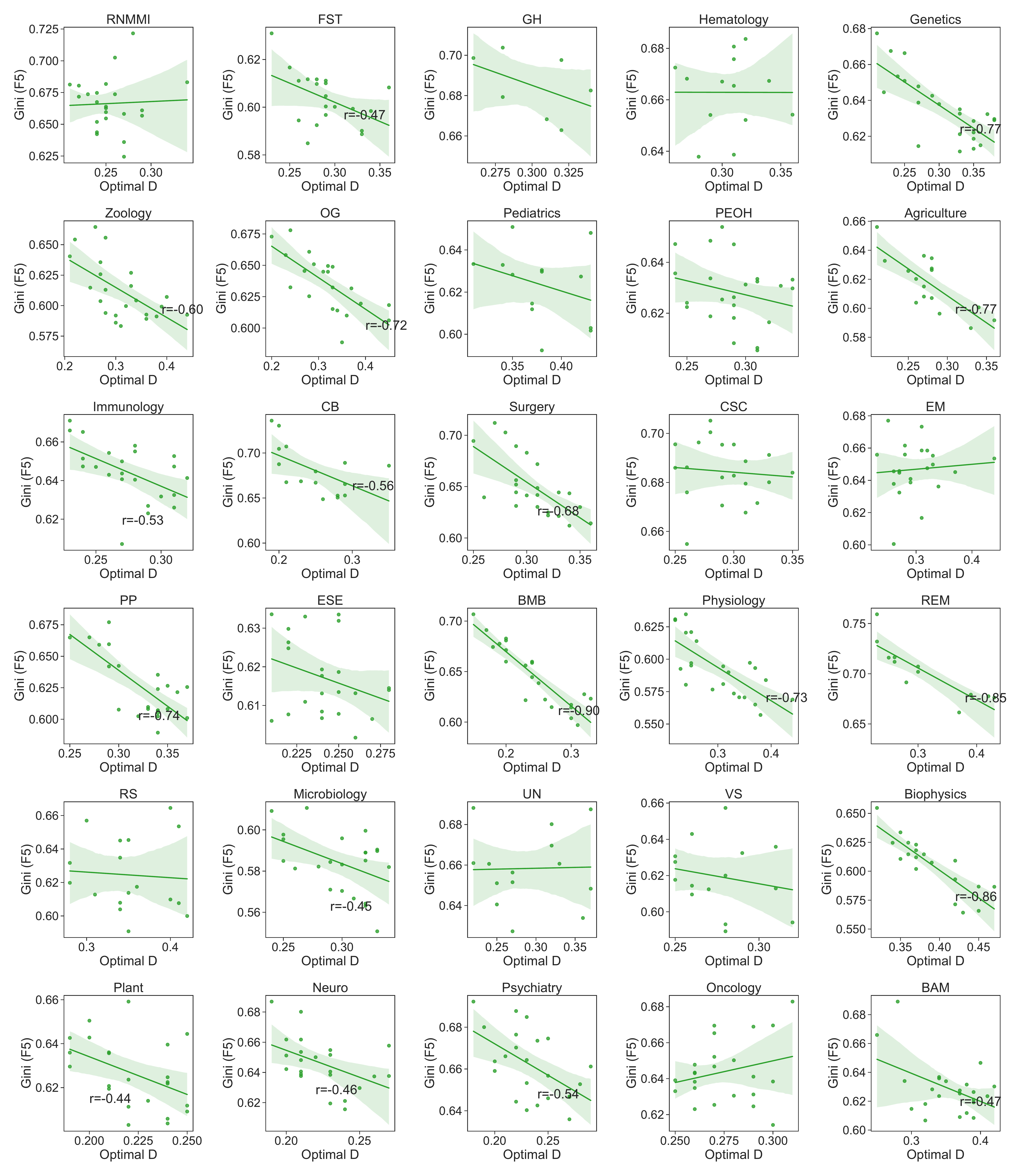}
    \caption{Correlation between impact ranking mobility and Gini coefficients in the Life Sciences \& Biomedicine. The Gini coefficient is calculated based on the cumulative number of citations authors have received in the first five years of their career.}
    \label{fig:Corr_life}
\end{figure*}

\begin{figure*}[ht!]
\centering
    \includegraphics[width=16cm]{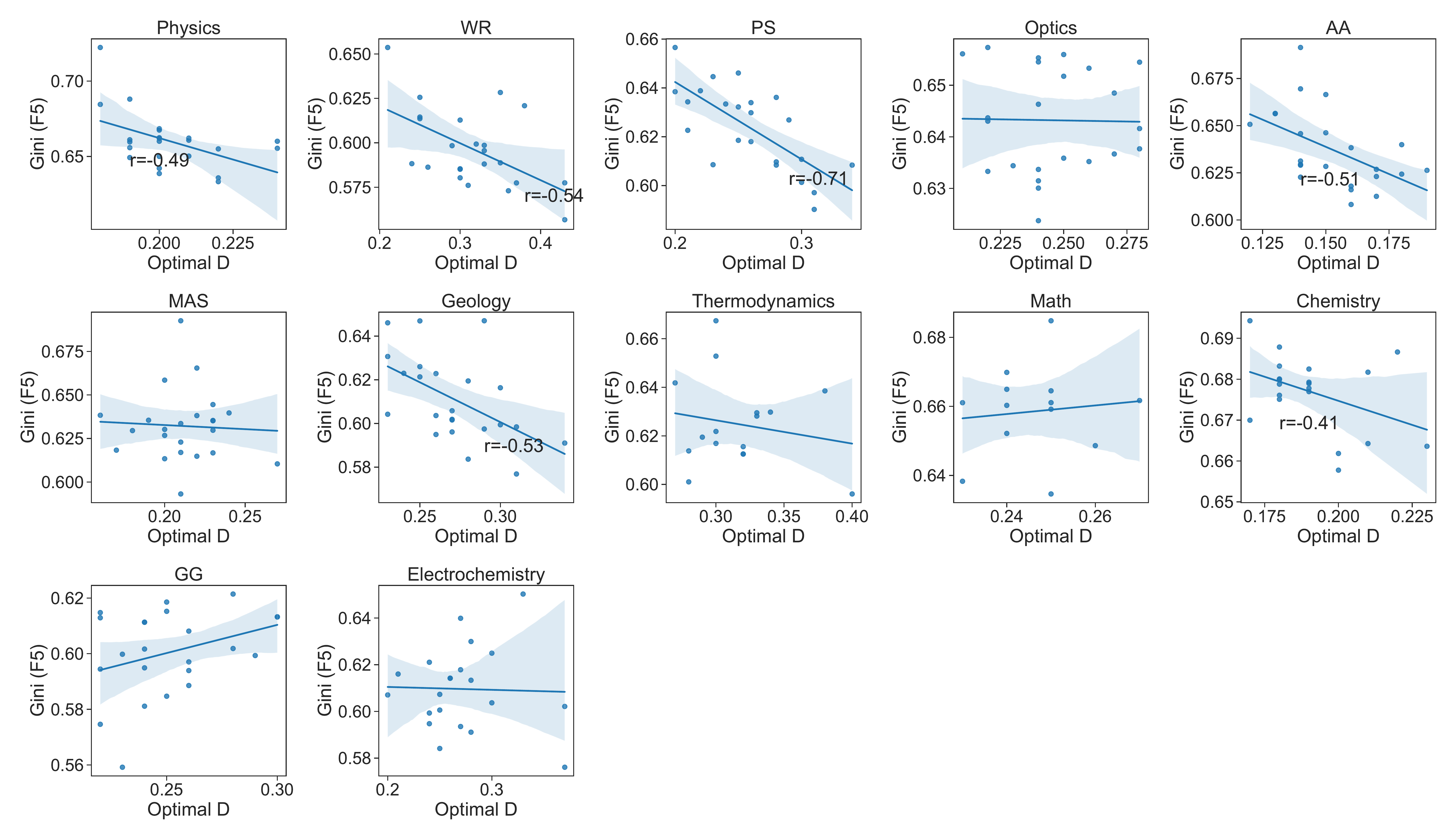}
    \caption{Correlation between mobility and Gini coefficients in the Physical Sciences. The Gini coefficient is calculated based on the cumulative number of citations authors have received in the first five years of their career.}
    \label{fig:Corr_physics}
\end{figure*}

\begin{figure*}[ht!]
\centering
    \includegraphics[width=16cm]{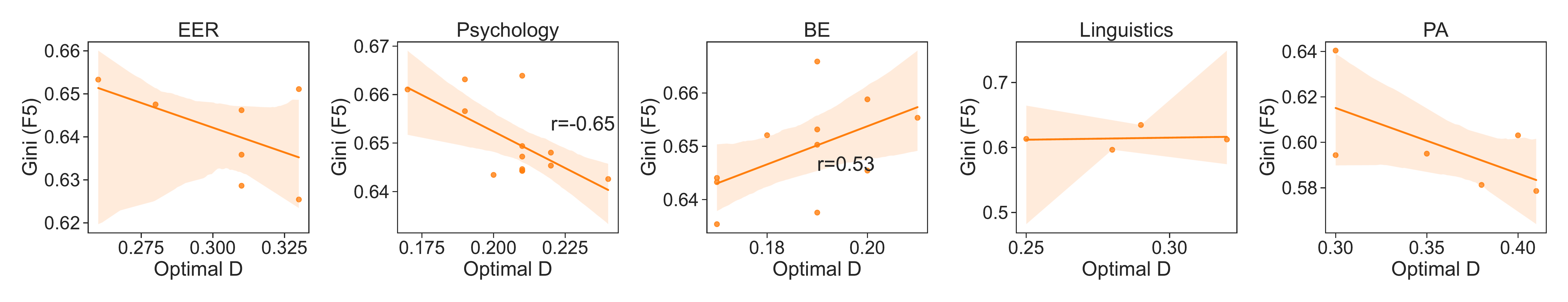}
    \caption{Correlation between mobility and Gini coefficients in the Social Sciences. The Gini coefficient is calculated based on the cumulative number of citations authors have received in the first five years of their career.}
    \label{fig:Corr_social}
\end{figure*}

\begin{figure*}[ht!]
\centering
    \includegraphics[width=16cm]{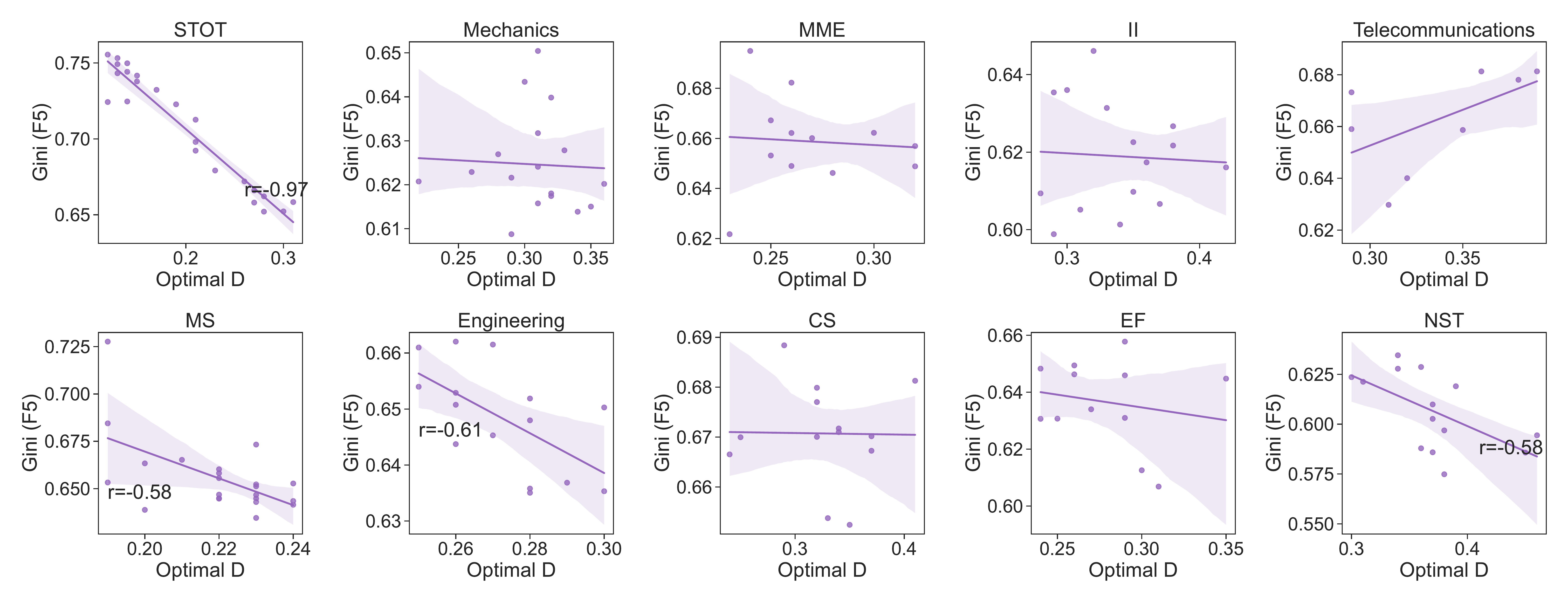}
    \caption{Correlation between mobility and Gini coefficients in Technology. The Gini coefficient is calculated based on the cumulative number of citations authors have received in the first five years of their career.}
    \label{fig:Corr_technology}
\end{figure*}

We validate these results by repeating the analysis at the level of individual disciplines (see Figs.~\ref{fig:AuthorMobility_life}-\ref{fig:AuthorMobility_technology} for trends in mobility and in Figs.~\ref{fig:AuthorGini_life}-\ref{fig:AuthorGini_technology} for trends in inequality). Fig.~\ref{fig:Evolution_Gini_Mobility}C summarizes the number of disciplines with statistically significant increasing trends of mobility and decreasing trends of inequality, confirming the generality of the trends observed above. In Fig.~\ref{fig:Correlation_Gini_Mobility}A we provide an overall representation of the negative relationship between mobility and inequality, by depicting each discipline and author cohort as a point. The negative correlation is still found at the level of four macro areas (Figs.~\ref{fig:Correlation_Gini_Mobility}B-E) and individual disciplines (Fig.~\ref{fig:Evolution_Gini_Mobility}C and Figs.~\ref{fig:Corr_life}-\ref{fig:Corr_technology}). 

\begin{figure*}[ht!]
\centering
    \includegraphics[width=16cm]{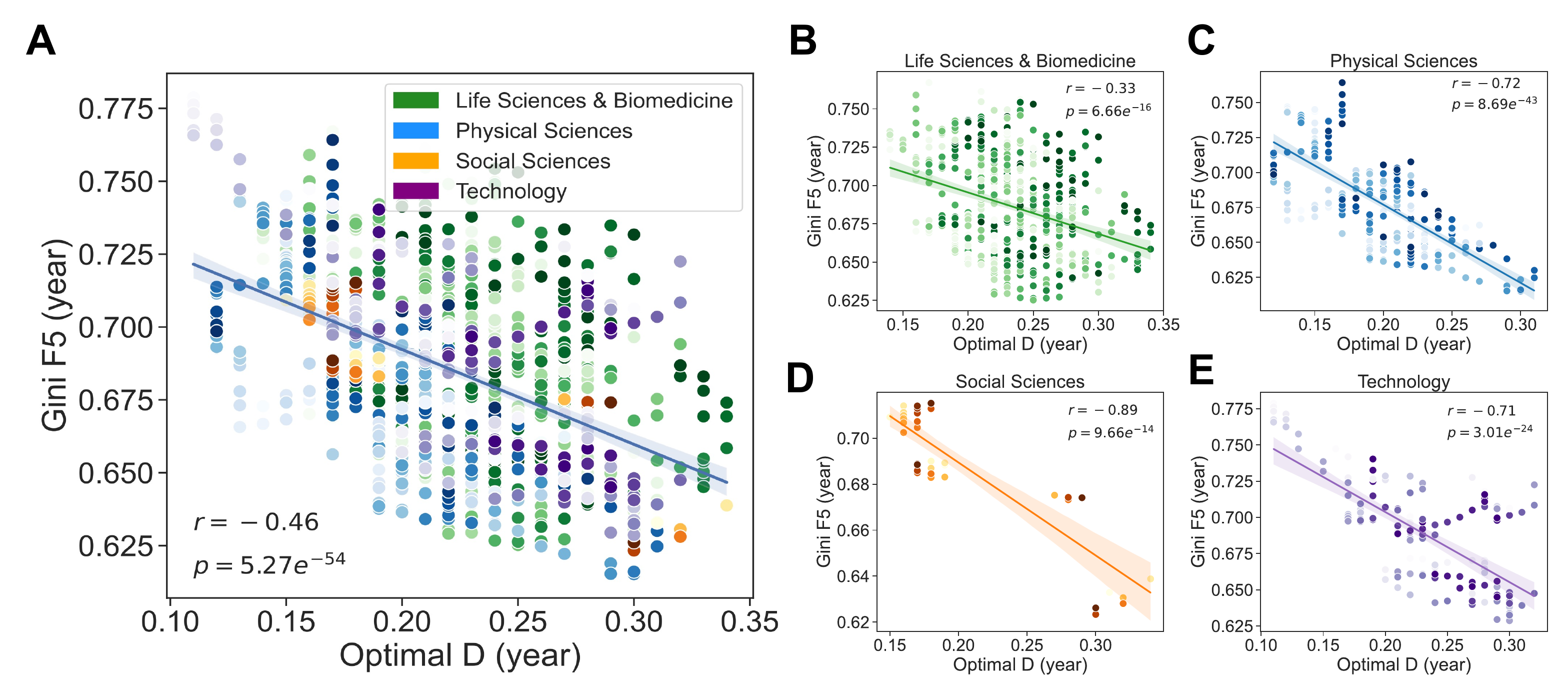}
    \caption{Correlation between mobility in impact rankings and inequality in impact distributions for authors across all disciplines (A), and for author cohorts in the Life science and Biomedicine (B), Physical Sciences (C), Social Sciences (D) and Technology (E). The Gini coefficient is calculated based on the cumulative number of citations authors have received within a 5-years time window (with starting year marked on the x-axis). In each subplot, each circle represents an author cohort in a given discipline. The shading of the circles, from light to dark, indicates the year in which a cohort of authors started their careers, from $1986$ to $2008$. The solid line and the shaded area indicate regression lines and $95\%$ confidence level intervals, respectively. Each regression has also been annotated with the corresponding Pearson correlation coefficient $r$ and its $p$-value.}
\label{fig:Global_Correlation_Gini_Mobility} 
\end{figure*}

\begin{figure*}[ht!]
\centering
    \includegraphics[width=16cm]{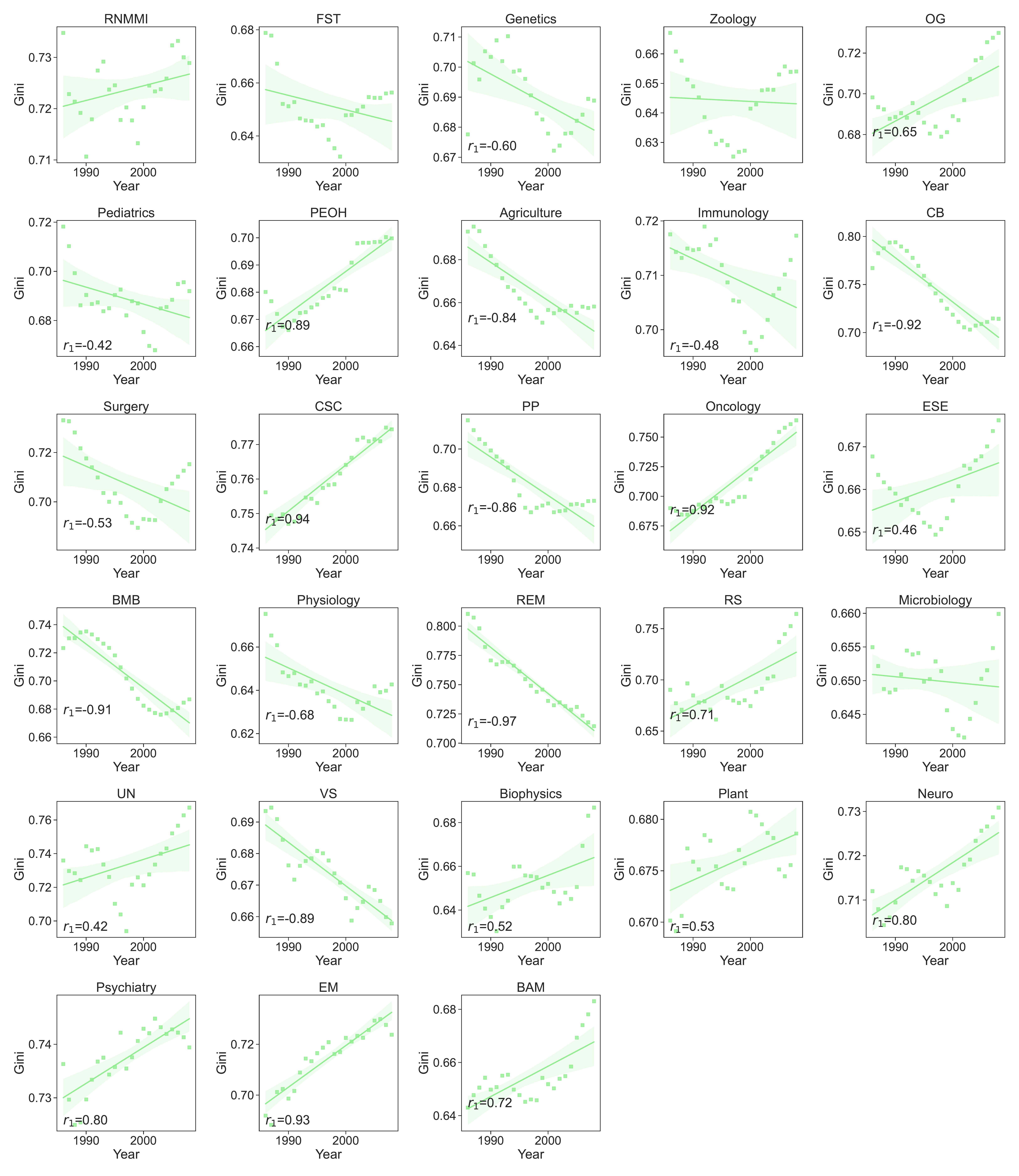}
    \caption{Gini coefficients of global author citation distributions in the Life Sciences \& Biomedicine from $1986$ to $2008$. The Gini coefficients are calculated based on the distribution of the cumulative number of citations obtained by authors within 5-year time windows (starting with the years shown on the x-axis). The disciplines with significant increasing/decreasing trends of mobility over time (with $p$-value < 0.1) are marked with Pearson correlation coefficient $r_{1}$.}
    \label{fig:Global_AuthorGini_life}
\end{figure*}

\begin{figure*}[ht!]
\centering
    \includegraphics[width=16cm]{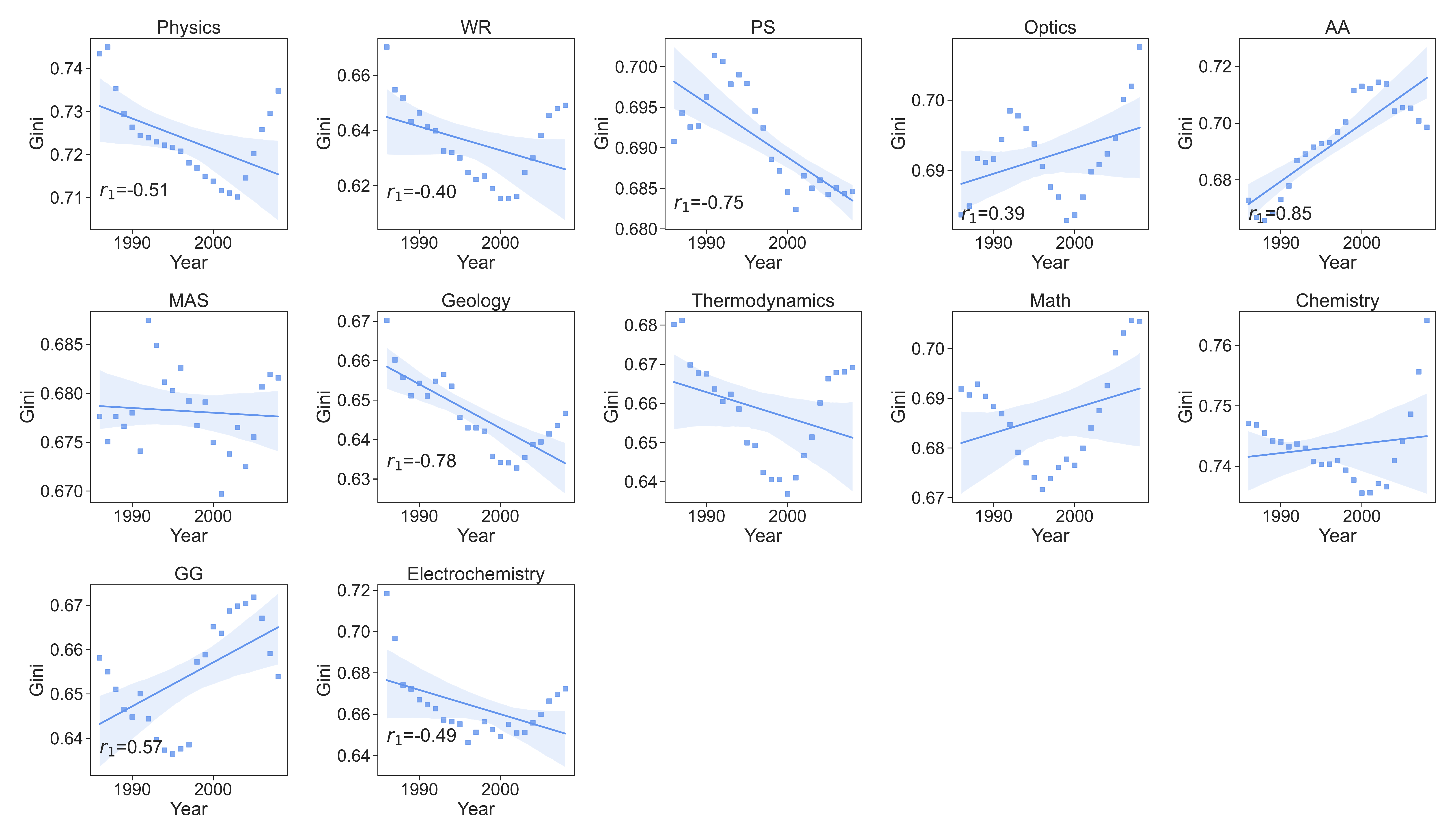}
    \caption{Gini coefficients of global author citation distributions in the Physical Sciences from $1986$ to $2008$. The Gini coefficients are calculated based on the distribution of the cumulative number of citations obtained by authors within 5-year time windows (starting with the years shown on the x-axis). The disciplines with significant increasing/decreasing trends of mobility over time (with $p$-value < 0.1) are marked with Pearson correlation coefficient $r_{1}$.}
    \label{fig:Global_AuthorGini_physics}
\end{figure*}

\begin{figure*}[ht!]
\centering
    \includegraphics[width=16cm]{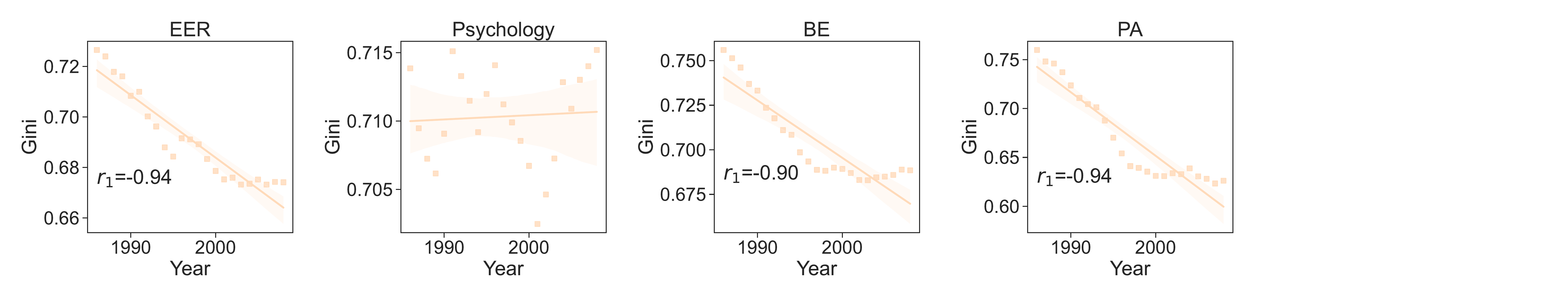}
    \caption{Gini coefficients of global author citation distributions in the Social Sciences from $1986$ to $2008$. The Gini coefficients are calculated based on the distribution of the cumulative number of citations obtained by authors within 5-year time windows (starting with the years shown on the x-axis). The disciplines with significant increasing/decreasing trends of mobility over time (with $p$-value < 0.1) are marked with Pearson correlation coefficient $r_{1}$.}
    \label{fig:Global_AuthorGini_social}
\end{figure*}

\begin{figure*}[ht!]
\centering
    \includegraphics[width=16cm]{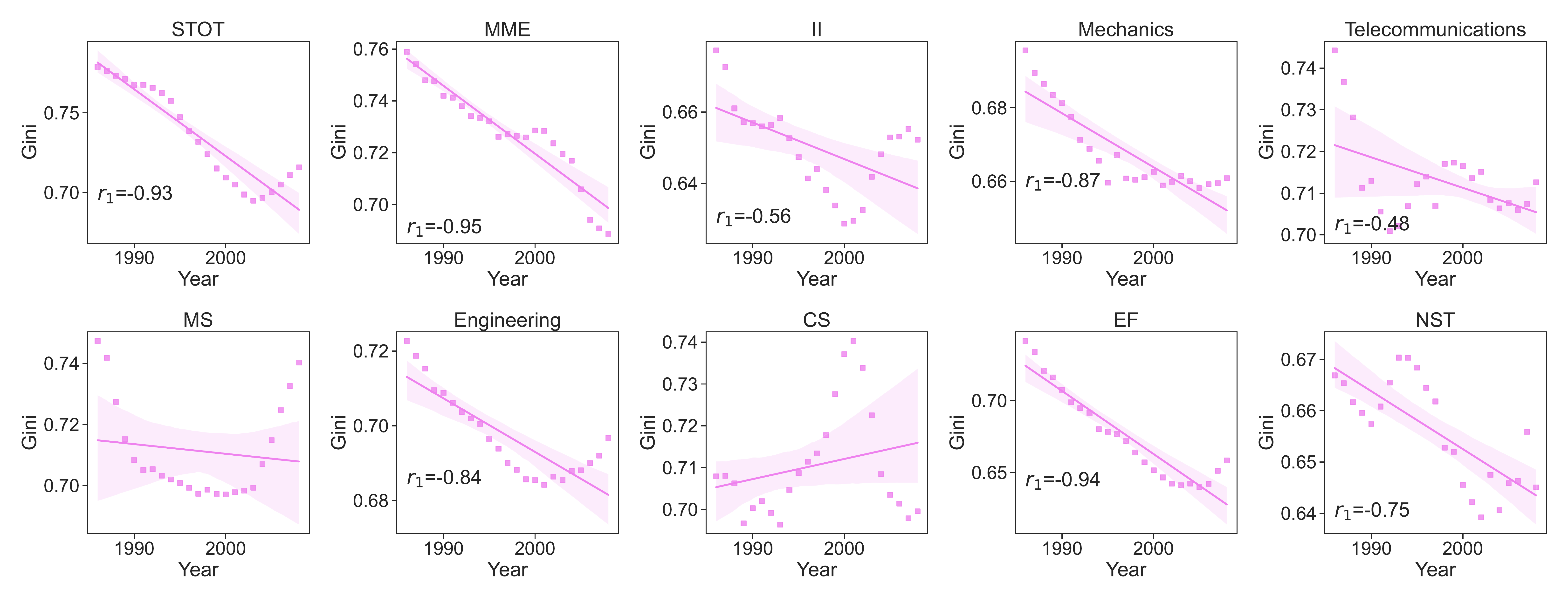}
    \caption{Gini coefficients of global author citation distributions in Technology from $1986$ to $2008$. The Gini coefficients are calculated based on the distribution of the cumulative number of citations obtained by authors within 5-year time windows (starting with the years shown on the x-axis). The disciplines with significant increasing/decreasing trends of mobility over time (with $p$-value < 0.1) are marked with Pearson correlation coefficient $r_{1}$.}
    \label{fig:Global_AuthorGini_technology}
\end{figure*}

Our results are obtained at the level of individual author cohorts. However, other studies~\cite{nielsen2021global} in the literature about academic impact inequality have focused on the entire population of active authors in a discipline at a given time, regardless of their academic seniority and/or career starting year. For consistency with this literature, we have repeated our analyses under the same settings, that is computing ranking mobility and inequality from the distributions of citations in a discipline at a given point in time. As shown in Fig.~\ref{fig:Global_Correlation_Gini_Mobility}, we still observe a negative relationship between ranking mobility and inequality, both on an aggregate level and at the level of macro areas. However, the temporal dynamics of both quantities display a more diverse set of behaviours. For instance, some disciplines are characterized by a U-shaped pattern in ranking inequality with a decreasing trend before the year 2000 and a rebound afterwards (Figs.~\ref{fig:Global_AuthorGini_life}-\ref{fig:Global_AuthorGini_technology}), which one could speculate being due to the rapid development of electronic publishing in the new millennium. 

\begin{table*}
\caption{Disciplines with highest or lowest mobility and inequality.}
\resizebox{\textwidth}{!}{\begin{tabular}{llll}
\hline
Top $5$ mobility & Bottom $5$ mobility & Top $5$  inequality & Bottom $5$ inequality  \\
\midrule
1. Biophysics                     & 1. Astronomy \& Astrophysics          & 1. Research \& Experimental Medicine  & 1. Microbiology \\
2. Pediatrics                     & 2. Science \& Technology Other Topics & 2. Cell Biology                       & 2. Physiology\\
3. Nuclear Science \& Technology  & 3. Business \& Economics              & 3. Science \& Technology Other Topics & 3. Water Resources \\
4. Respiratory System             & 4. Chemistry                          & 4. Gastroenterology \& Hepatology     & 4. Geochemistry \& Geophysics\\
5. Public Administration          & 5. Physics                            & 5. Cardiovascular System \& Cardiology& 5. Food Science \& Technology\\
\bottomrule
\end{tabular}}
\label{tab:TobBottomDisciplines}
\end{table*}

The observed differences in mobility and inequality across disciplines may reflect the different underlying opportunities for researchers entering an academic discipline. To rank the disciplines in terms of their overall impact mobility and inequality, we first compute a single discipline-specific value of the parameter $D$ by calibrating our random walk model on the transition matrices of a given discipline for all years in our analysis. We do that by finding the value of the parameter which minimizes the sum of Frobenius norms of the differences between empirical transition matrices and those generated by the model. Similarly, we can characterize the overall inequality of a discipline by computing its average annual Gini coefficient. This naturally leads to a ranking of the disciplines in terms of their overall ranking mobility and inequality. In Table~\ref{tab:TobBottomDisciplines} we report the top/bottom 5 disciplines in both dimensions. We find that the discipline characterised by the highest (lowest) overall mobility is Biophysics (Astronomy \& Astrophysics), whereas the discipline characterised by the highest (lowest) overall inequality is Research \& Experimental Medicine (Microbiology).

\section*{Discussion}
In this paper, we have studied the careers of a large pool of authors across $57$ disciplines, focusing on impact ranking mobility and inequality at the level of cohorts, that is we only compare authors with peers starting their careers in the same year. We document the existence of a statistically significant negative relationship between these quantities. This is reminiscent of a similar phenomenon observed in the social sciences, where metrics of wealth inequality are often shown to be negatively correlated with metrics of social mobility~\cite{wilkinson2006impact}. From the perspective of this analogy, one could be tempted to identify ranking mobility in a discipline as the academic equivalent of social mobility in a country, and citations as the academic equivalent of wealth. Although academic impact is certainly a multifaceted concept, it is in fact undeniable that citations and citation-based indicators have become the currency of academic progression in a number of countries and academic systems~\cite{moher2018assessing}, sometimes leading to unintended consequences on citation behaviours and patterns~\cite{li2019reciprocity,baccini2019citation}. To push this analogy even further, a number of studies have shown that academic impact, as measured via citations, is partially inherited from mentors and/or senior collaborators~\cite{sekara2018chaperone,li2019early,kelty2023don,yadav2023does}, quite similarly to family wealth which is passed on through generations. We wish to stress that here we are not endorsing an uncritical use of citations and citation-based indicators as a measure of academic impact. Quite the contrary, in light of our results it is even clearer that such metrics should be properly contextualized.

When monitoring the temporal evolution of ranking mobility and inequality, we find that -- in the majority of disciplines -- the former has been on the rise while the latter has decreased over extended periods of time. This is in contrast with common perception, as supported by the extensive literature on the Matthew effect in academia~\cite{merton1968matthew,petersen2014reputation,bol2018matthew}. However, our units of analysis are different from those typically considered in that literature. While the latter compares each author with all other authors in their discipline that are active at the same time, regardless of their seniority, we compare each author with those that started their careers in the same year. In other words, our measures of ranking mobility and inequality are based on conditional probabilities over cohorts rather than the entire population of active authors. In fact, when repeating our analyses at the level of the entire population we find increasing inequality (after the year 2000) in a plurality of disciplines. Moreover, intra-cohort mobility mostly takes place in the middle strata of the cohort, with the lowest and highest deciles being characterized by a notable lack of mobility (as well captured by the discrepancies with respect to our random walk model of mobility). Coming back to the previous analogy with the social sciences, this is reminiscent of a long-standing observation (first made by Pareto~\cite{Pareto}) that most social mobility takes place in the middle classes~\cite{bardoscia2013social}.

Put together, the above results suggest that over time it has become easier for new authors to climb the ranks of their own cohorts, while at the same time they may experience a higher inequality at the level of entire populations. A possible explanation for such a seeming paradox may lie in the increasing volume of publications~\cite{bornmann2021growth} and citations, which effectively reduces over time the intrinsic value of a citation~\cite{siler2022games}.

\section*{Methods}
\subsection*{Dataset}
The bibliometric data for this study are provided by the Web of Science (WoS). Publications are classified into $153$ disciplines, which constitute a subject categorization scheme that is shared by all Web of Science product databases. One publication can be assigned to more than one discipline. Disciplines are further grouped into five broad macro areas: 1) Arts \& Humanities; 2) Life Sciences \& Biomedicine; 3) Social Sciences; 4) Physical Sciences; 5) Technology. For the analysis, we select several of the largest disciplines within each macro area: $30$ disciplines in Life Sciences, $12$ in Physical Sciences, $5$ in Social sciences and $8$ in Technology. The disciplines in Arts \& Humanities are excluded from our analysis because a large portion of articles in this macro area do not receive any citations after publication. Note that the publications with more than $20$ listed authors are not considered in our analysis.   

\subsection*{Name disambiguation algorithm}
The main focus of our work is to examine the mobility of authors' scientific impact rankings in their early career. Therefore, adequate disambiguation of author names in bibliometric databases is pivotal for any reliable analysis at the author level. The WoS dataset does not maintain unique author identifiers. To associate an author to all of their publications, we apply a state-of-the-art algorithm proposed by Caron and van Eck~\cite{caron2014large} to disambiguate all names of authors starting their academic career after $1986$ in the entire WoS database. This approach has been validated by Tekles and Bornmann, who showed that it outperforms four other unsupervised disambiguation methods~\cite{tekles2020author} in large-scale bibliometric analysis. Specifically, this method quantifies the similarity between two author mentions using rule-based scoring and clustering. A set of crtieria that rely on several paper-level and author-level attributes have been considered, including ORCID identifiers, names, affiliations, email addresses, coauthors, grant numbers, subject categories, journals, self-citations, bibliographic coupling, and co-citations. Each criterion is assigned a specific score, and the scores of all matching criteria are summed to give an overall similarity score for the two author mentions. The higher the similarity score of the two author mentions, the more likely they are to be considered the same author. The algorithm's performance has been shown to exceed 90\% both in terms of of precision and recall. 

\bibliographystyle{naturemag}
\bibliography{scibib}

\section*{Data availability}
The data used in the study are fully available with a subscription to Web of Science. All other data are included in the manuscript and/or SI Appendix.


\section*{Acknowledgements}
Y.S. and G.L. acknowledge support from a Leverhulme Trust research project grant (RPG-2021-282).

\section*{Author contributions}
Y.S., F.C. and G.L. conceived and designed research; Y.S. performed research; Y.S., F.C. and G.L. analyzed data; Y.S., F.C. and G.L. wrote and edited the paper.

\section*{Competing interests}
The authors declare that they have no competing interests.

\clearpage

\beginsupplement

\begin{titlepage}
    \begin{center}
        \huge
        Supplementary Information for:\\
        \huge
         Ranking mobility and impact inequality in early academic careers\\
        \vspace{0.5cm}
        \Large
        Ye Sun$^{1}$, Fabio Caccioli$^{1,2,3}$, Giacomo Livan$^{1,2,\ast}$\\
        
        \vspace{0.5cm}
        \large
        $^{1}$Department of Computer Science, University College London, 66-72 Gower Street, London WC1E 6EA, United Kingdom\\
        $^{2}$London School of Economics and Political Science, Systemic Risk Centre, London WC2A 2AE, United Kingdom\\
        $^{3}$London Mathematical Laboratory, United Kingdom\\
        \vspace{0.5cm}
        \large
        $^{\ast}$Corresponding author. Email: g.livan@ucl.ac.uk.
        \end{center}
\end{titlepage}


\clearpage

\begin{table*}[bt!]
\centering
\small
\renewcommand\arraystretch{0.9}  
\caption{\textbf{The four macro areas considered in our analysis, and the list of $57$ disciplines within them.}}
\begin{tabular}{p{3cm}p{6.8cm}p{2.5cm}p{1.5cm}p{0.9cm}p{0.9cm}}
\hline
\textbf{Macro Area}               & \textbf{Discipline}        & \textbf{Abbreviation}  & \textbf{$N_p$}  & \textbf{$\bar{D}$} & \textbf{$\bar{G}$} \\ \hline
Life Sciences \& Biomedicine       & 1. Radiology, Nuclear Medicine \& Medical Imaging   & RNMMI       & 751,906  & 0.25 & 0.67\\
                                   & 2. Food Science \& Technology          & FST         & 502,764  & 0.29 & 0.61 \\
                                   & 3. Gastroenterology \& Hepatology      & GH          & 744,741  & 0.30 & 0.69 \\
                                   & 4. Hematology                          & Hematology  & 683,544  & 0.30 & 0.67\\
                                   & 5. Genetics \& Heredity                & Genetics    & 661,292  & 0.30 & 0.63\\
                                   & 6. Zoology                             & Zoology     & 422,719  & 0.31 & 0.61\\
                                   & 7. Obstetrics \& Gynecology            & OG          & 481,532  & 0.32 & 0.64\\
                                   & 8. Pediatrics                          & Pediatrics  & 603,328  & 0.38 & 0.63\\
                                   & 9. Public, Environmental \& Occupational Health   & PEOH  & 811,069  & 0.29 & 0.63\\
                                   & 10. Agriculture                        & Agriculture      & 847,230  & 0.28 & 0.63\\
                                   & 11. Immunology                         & Immunology       & 956,010  & 0.27 & 0.64\\
                                   & 12. Cell Biology                       & CB        & 1,200,052  & 0.25 & 0.71\\
                                   & 13. Surgery                            & Surgery   & 1,294,252  & 0.30 & 0.65\\
                                   & 14. Cardiovascular System \& Cardiology & CSC      & 1,412,352  & 0.29  & 0.69\\
                                   & 15. Endocrinology \& Metabolism        & EM     & 764,274       & 0.29  & 0.65\\
                                   & 16. Pharmacology \& Pharmacy           & PP     & 1,462,975     & 0.32  & 0.63\\
                                   & 17. Environmental Sciences \& Ecology  & ESE    & 1,400,625     & 0.24  & 0.62\\
                                   & 18. Biochemistry \& Molecular Biology  & BMB    & 2,403,986     & 0.24  & 0.65\\
                                   & 19. Physiology                         & Physiology  & 432,717  & 0.30  & 0.59\\
                                   & 20. Research \& Experimental Medicine	& REM     & 728,682  & 0.30  & 0.73 \\
                                   & 21. Respiratory System                 & RS      & 493,372  & 0.35  & 0.63\\
                                   & 22. Microbiology                       & Microbiology  & 578,183    & 0.30 & 0.58\\
                                   & 23. Urology \& Nephrology	            & UN     & 525,344  & 0.28   & 0.67\\
                                   & 24. Veterinary Sciences                & VS     & 468,134  & 0.28   & 0.63\\
                                   & 25. Biophysics	                        & Biophysics  & 515,504  & 0.39   & 0.61\\
                                   & 26. Plant Sciences	                    & Plant     & 657,073    & 0.22   & 0.63\\
                                   & 27. Neurosciences \& Neurology         & Neuro     & 2,171,568  & 0.22   & 0.65\\
                                   & 28. Psychiatry                         & Psychiatry  & 734,915  & 0.23   & 0.66\\
                                   & 29. Oncology	                        & Oncology    & 1,516,450  & 0.27 & 0.64\\
                                   & 30. Biotechnology \& Applied Microbiology & BAM    & 663,092    & 0.35  & 0.63\\
                                   \hline    
Physical Sciences                  & 1. Physics                             & Physics   & 3,492,227   & 0.20  & 0.66  \\
                                   & 2. Water Resources                     & WR        & 280,079     & 0.31  & 0.60  \\
                                   & 3. Polymer Science                     & PS        & 469,508     & 0.26  & 0.62  \\
                                   & 4. Optics                              & Optics    & 588,541     & 0.25  & 0.62  \\
                                   & 5. Astronomy \& Astrophysics           & AA        & 514,016     & 0.15  & 0.64   \\
                                   & 6. Meteorology \& Atmospheric Sciences & MAS       & 271,045     & 0.21  & 0.63   \\
                                   & 7. Geology                             & Geology   & 538,634     & 0.27  & 0.61        \\
                                   & 8. Thermodynamics                      & Thermodynamics  & 252,127 & 0.31  & 0.63       \\
                                   & 9. Mathematics                         & Math      & 1,252,614  & 0.25 & 0.66        \\
                                   & 10. Chemistry                          & Chemistry & 4,649,221  & 0.19 & 0.68       \\
                                   & 11. Geochemistry \& Geophysics         & GG        & 267,929    & 0.25 & 0.60        \\
                                   & 12. Electrochemistry                   & Electrochemistry  &  260,391  & 0.27 &  0.62      \\
                                   \hline
Social Sciences                    & 1. Education \& Educational Research   & EER       &  382,374    &  0.30 & 0.66    \\
                                   & 2. Psychology                          & Psychology&  1,213,754  &  0.21 & 0.66       \\
                                   & 3. Business \& Economics               & BE        &  909,917    &  0.19 &  0.68     \\
                                   & 4. Linguistics                         & Linguistics & 109,417   &  0.29 &  0.64      \\
                                   & 5. Public Administration               & PA         & 116,976    &  0.35 &  0.65      \\    
                                   \hline
Technology                         & 1. Science \& Technology Other Topics  & STOT       & 1,678,118    &  0.19  &  0.71   \\
                                   & 2. Mechanics                           & Mechanics  & 429,713      &  0.31  &  0.63    \\
                                   & 3. Metallurgy \& Metallurgical Engineering & MME    & 401,500      &  0.27  &  0.68    \\
                                   & 4. Instruments \& Instrumentation       & II        & 379,571      &  0.34  &  0.62    \\
                                   & 5. Telecommunications                   & Telecommunications &  342,441  & 0.33 & 0.68  \\ 
                                   & 6. Materials Science                    & MS         &  2,462,909   &  0.22  &  0.66   \\
                                   & 7. Engineering                          & Engineering&  3,686,638   &  0.27  &  0.66   \\
                                   & 8. Computer Science                     & CS         &  1,191,832   &  0.33  &  0.67   \\
                                   & 9. Energy \& Fuels                      & EF         &  487,127     &  0.28  &  0.66   \\
                                   & 10. Nuclear Science \& Technology       & NST        &  275,289     &  0.37  &  0.61   \\
\bottomrule
\end{tabular}
\caption*{$N_{p}$ represents the total number of papers published by authors in each discipline. $\bar{D}$ indicates the overall optimal impact mobility, calculated by minimising the sum of Frobenius norms of the differences between empirical transition matrices and those obtained from the random walk mode throughout the whole period for each discipline. $\bar{G}$ is the average annual Gini coefficient over time for each discipline.}
\label{tab:ScoreRules}
\end{table*}

\end{document}